\documentclass{aastex631}

\shorttitle{Mass Transfer in IT Lib}
\shortauthors{Wysocki et al.}
\graphicspath{{./}{figures/}}

\begin{document}

\title{Mass Transfer as an Explanation for the Lifetime-Travel Time Discrepancy in IT Librae}

\author[0000-0003-0392-1094]{Peter Wysocki}
\affiliation{Center for High Angular Resolution Astronomy, Department of Physics and Astronomy, Georgia State University, P.O. Box 5060, Atlanta, GA 30302-5060, USA}

\author[0000-0001-8537-3583]{Douglas Gies}
\affiliation{Center for High Angular Resolution Astronomy, Department of Physics and Astronomy, Georgia State University, P.O. Box 5060, Atlanta, GA 30302-5060, USA}

\author[0000-0003-2075-5227]{Katherine Shepard}
\affiliation{Center for High Angular Resolution Astronomy, Department of Physics and Astronomy, Georgia State University, P.O. Box 5060, Atlanta, GA 30302-5060, USA}

\author[0000-0002-9903-9911]{Kathryn Lester}
\affiliation{NASA Ames Research Center, Moffett Field, CA 94035, USA}

\author[0000-0001-9647-2886]{Jerome Orosz}
\affiliation{Department of Astronomy, San Diego State University, 5500 Campanile Drive, San Diego, CA 92182, USA}



\begin{abstract}

The eclipsing binary IT Librae is an unusual system of two B-type stars
that is situated about 1 kpc above the galactic plane. The binary was
probably ejected from its birthplace in the disk, but the implied
time-of-flight to its current location exceeds the evolutionary
lifetime of the primary star.  Here we present a study of new
high dispersion spectroscopy and an exquisite light curve from the
Kepler K2 mission in order to determine the system properties and
resolve the timescale discrepancy.  We derive a revised spectroscopic
orbit from radial velocity measurements and determine the component
effective temperatures through comparison of reconstructed and
model spectra ($T_1 = 23.8 \pm 1.8$ kK, $T_2 = 13.7 \pm 2.5$ kK).
We use the Eclipsing Light Curve (ELC) code to model the K2 light curve,
and from the inclination of the fit, we derive the component
masses ($M_1 = 9.6 \pm 0.6 M_\odot$, $M_2 = 4.2 \pm 0.2 M_\odot$) and
mean radii ($R_1 = 6.06 \pm 0.16 R_\odot$, $R_2 = 5.38 \pm 0.14 R_\odot$).
The secondary star is overluminous for its mass and appears to fill
its Roche lobe.  This indicates that IT~Librae is a post-mass transfer
system in which the current secondary was the mass donor star.
The current primary star was rejuvenated by mass accretion, and its
evolutionary age corresponds to the time since the mass transfer stage.
Consequently, the true age of the binary is larger than the ejection
time-of-flight, thus resolving the timescale discrepancy.

\end{abstract}



\section{Introduction} \label{sec:intro}
The Milky Way galaxy hosts a unique population of hot stars known as High-Galactic Latitude B Stars. This group is composed of faint blue stars at high galactic latitudes, often with high peculiar velocities. They were first discovered in surveys looking for white dwarfs \citep{1947ApJ...105...85H}. Many of these blue stars are not actually main sequence B stars, but other objects masquerading as them. Most are white dwarfs or OB subdwarfs (sdOB), which is a group of blue stars primarily composed of extreme blue horizontal branch stars, but also includes post-planetary nebula stars approaching the top of the white dwarf sequence \citep{1986ApJS...61..305G}. There is also a very rare population of He-rich sdOB stars, thought to be formed through white dwarf mergers \citep{2004ApSS.291..253A}. However, not all of these high latitude B stars are old evolved stars; some are young main sequence (MS) B stars \citep{1992QJRAS..33..325K}. This population is unique in that the B stars are far outside of the galactic plane where they were born. B stars have masses in the range of 2 - 20 $M_{\odot}$, so their lifetimes will be significantly shorter than that of most other stars. This creates a problem for how they can exist in the galactic halo. 

One theory is that the B stars formed in high velocity gas clouds (HVCs) in the galactic halo. HVCs have average masses of hundreds to thousands of solar masses, and cores averaging one to ten solar masses \citep{1993ASPC...45...11V}. However, there seems to be no unambiguous evidence for star formation in these clouds \citep{Willman_2002, Simon_2002}. Despite a few examples to the contrary \citep{Ivezi__1997}, the evidence overall points to the star formation rate being very low in HVCs, so it is doubtful halo B-stars formed in situ. Therefore, because few, if any, B stars form in the galactic halo, there must be some dynamical process to get them there relatively quickly. 

Another theory is that the B stars were ejected from the galactic plane. Such high velocity O- and B-type runaway stars attain their large speed in two ways \citep{2001AA...365...49H}: the binary-supernova scenario (BSS) and the dynamical ejection scenario (DES).  In the binary-supernova scenario, a massive binary begins with a primary star that is more massive than the secondary and it evolves first. It explodes in a supernova and launches the secondary star with a high peculiar velocity comparable to the orbital speed. In many cases, the stellar remnant will remain bound to the runaway secondary as either a neutron star or a black hole. In other cases, the secondary star instead may become unbound and become a single runaway star. Ejected binaries may originate from a triple consisting of a binary system orbiting a single star more massive than either binary component. The single star will evolve first and explode, leaving the orbiting binary unbound. An analysis of these systems and their typical ejection velocities is described by \cite{2019MNRAS.487.3178G}. In the dynamical ejection scenario, runaway stars are created through gravitational interactions between stars. Close encounters between single or binary stars with binaries may result in the hardening (shrinkage of orbital separation) of one binary and the ejection of single or sometimes binary stars \citep{1990AJ.....99..608L}. These ejection velocities can be quite high, up to 200 km~s$^{-1}$ (or even higher in rare cases), and so can easily create runaway single and binary stars \citep{2012ApJ...751..133P}.

IT Librae (HD 138503) is located at (343$^{\circ}$, +24$^{\circ}$) in J2000 galactic coordinates and is a member of a rare group of high latitude B stars that are also eclipsing binaries \citep{1999IBVS.4659....1K}. It is located about 1 kpc above the galactic plane \citep{2003PASP..115...49M}, which, as mentioned above, is very unusual for one B star, let alone two. The OB star scale height is estimated to only be about 25-65 pc, placing them in the ``extreme thin disk'' \citep{2000AJ....120..314R}. Only two other high latitude B star eclipsing binaries are known at distances greater than 1 kpc from the disk, V Tuc \citep{1918ApJ....48..310D} at galactic coordinates (303$^{\circ}$, $-72^{\circ}$) and DO Peg \citep{1935AN....255..401H} at galactic coordinates (67$^{\circ}$, $-38^{\circ}$). IT Librae is the brightest and the most easily observed member of the rare group. One particularly important question about high latitude B stars is how to address the tension between their ages and the travel time it takes to get to their current location. B stars typically only live for about 10 -- 300 million years, and with distances for some of these halo B stars placing them over a kiloparsec away from the galactic plane, it can take tens of millions of years to get to their current location. Therefore, depending on the mass of the B stars, the predicted travel times could be very similar to or even exceed the lifetimes of the stars. See \citet{2009ApJ...698.1330P} for a list of these too young stars in the halo.

The first detailed work on IT Lib was published by \cite{2003PASP..115...49M}. He used Hipparcos photometric data and high resolution spectra from the Sandiford Echelle Spectrograph at McDonald Observatory (7 spectra collected from 1999 to 2002) to determine the orbital parameters of the system, plus the temperatures of the components. He found masses of $9.8 \pm 0.7 M_{\odot}$ for the primary and $4.6 \pm 0.3 M_{\odot}$ for the secondary, and temperatures of 22,000 K for the primary and 13,620 K for the secondary. He also calculated IT Lib's trajectory and determined that it took about 33 million years for it to reach its current position, much longer than the evolutionary lifetime of about 22 million years for the $10 M_\odot$ primary \citep{doi:10.1146/annurev.aa.05.090167.003035}. Martin argued that the differences between the two timescales might be avoided with a more complete consideration of the uncertainties. Indeed, a recent discovery and analysis of 12 new runaway MS stars \citep{2021AA...645A.108R} shows they all have evolutionary lifetimes longer than their time-of-flight, as is expected for normal runaway stars.

Here we return to the issue of the origin of IT Lib through an analysis of new spectroscopy and photometry. Instead of a Hipparcos light curve, we use a more sampled, higher quality Kepler light curve, along with a detailed modeling code. We have collected 19 new (and one archival) high dispersion and high S/N spectra for determination of the radial velocity curve and physical properties. With these new datasets we can obtain a better understanding of IT Lib and its age and time-of-flight discrepancy. We can both confirm the existence of this discrepancy, and look for clues that could explain how the primary star has lived so long. Sections 2 and 4 discuss the Kepler K2 photometry, and Sections 3 and 5 describe the spectroscopic results. We consider the distance and time-of-flight estimates in Sections 6 and 7. We conclude with a discussion of the evolution of the binary in Sections 8 and 9.

\section{Period and Epoch Determination} \label{sec:Per}
\cite{2003PASP..115...49M} adopted an orbital period of $P=2.267460$ d from Hipparcos photometry. Fortunately IT~Lib was included in a Kepler K2 field of view (Campaign 15), so we can revisit the orbital ephemeris with the Kepler photometry. The long cadence (29.4 minute exposure time) Kepler light curve from the K2 mission was downloaded from the MAST Database\footnote{https://mast.stsci.edu/portal/Mashup/Clients/Mast/Portal.html}. It contains relative photometry covering the period from 2017 August 23 to 2017 November 19. This includes 39 primary eclipses, so 38 period timings are gained by fitting a parabola to the photometry obtained near eclipse minima, and finding the elapsed time between each primary minimum. However, these observations cover a relatively short time frame. In order to extend the baseline and substantially improve our accuracy, we also sought previous eclipse measurements. We found much older eclipse timings from 2005 June 6 \citep{2006IBVS.5690....1K}, and from average times centered on 2001 January 8 for the All Sky Automated Survey \citep{1997AcA....47..467P} and centered on 1991 January 6 for Hipparcos \citep{1997AA...323L..49P}. 

We found additional photometry from the Digital Access to a Sky Century at Harvard project (DASCH), an ongoing effort to digitize 100 years of sky surveys from photographic plates \citep{Laycock2010}.  We retrieved the photographic magnitudes from the DASCH web site\footnote{http://dasch.rc.fas.harvard.edu/lightcurve.php}, and we divided the time series into three bins: 1888 -- 1926 (597 measurements), 1927 -- 1953 (973 measurements), and 1970 -- 1989 (129 measurements). In each case, we transformed the observed times to orbital phase using the provisional period and epoch from the K2 data, and then we determined the best fit offset to the K2 light curve to find the time of primary eclipse minimum closest to the average time.  The derived primary eclipse times
are presented in Table \ref{tab:Min} and give us over 17000 orbital cycles from which to draw an accurate period. 

We made a linear fit of primary eclipse Barycentric Julian Date (BJD) vs. Cycle Number (E) to find a period of $P = 2.2674456 \pm 0.0000003$ days, and an epoch (time of Hipparocs epoch) of BJD $2448262.6772 \pm 0.0014$.  This period also agrees with that found recently by the SuperWASP Variable Stars citizen science project \citep{2021RNAAS...5..228M}. They report a period of 195,907.75 seconds (2.267451 days), which is consistent with our findings within errors. 

\begin{figure}[h!]
    \begin{center}
        \includegraphics[scale=0.55]{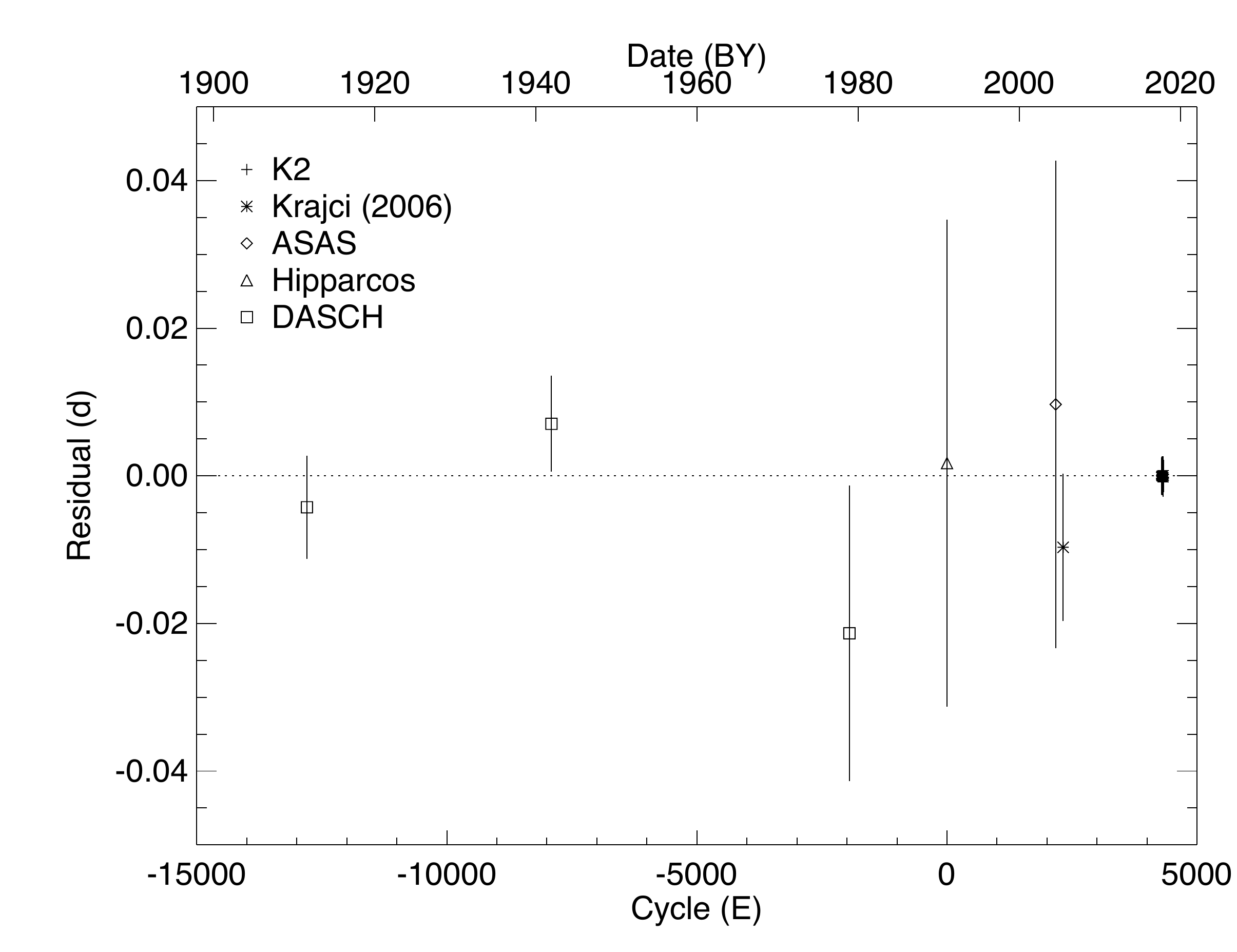}
    \caption{Residuals from the linear empheris fit of the primary eclipse time versus Cycle Number (E) for IT Lib.}
    \label{fig:PPlot}
    \end{center}
\end{figure}

Figure \ref{fig:PPlot} shows the residuals from the fit for our determination of the period. The cycle number gives the number of times a full period has been completed from a starting epoch, and because we know the period well from the Kepler light curve, the cycle numbers do not carry any uncertainty. We adopt our new ephemeris in the following sections.

\begin{deluxetable*}{ccc}
\tablenum{1}
\tablecaption{Times of Primary Minima \label{tab:Min}}
\tablewidth{0pt}
\tablehead{
\colhead{Time} & \colhead{Orbit Number} & \colhead{Source} \\
\colhead{(BJD -- 2400000)}  & \colhead{$E$} & \colhead{}
}
\decimalcolnumbers
\startdata
19246.1716 & $-12797$ & DASCH \\ 
30329.4570 & $-7909$ & DASCH \\
43838.8695 & $-1951$ & DASCH \\
48262.6789 & 0 &     Hipparcos \\
53189.8462 & 2173 & ASAS \\
53527.6762 & 2322 &     Krajci (2006) \\
57990.0194 & 4290 &     K2 \\
57992.2860 & 4291 &     K2 \\
57994.5532 & 4292 &     K2 \\
57996.8212 & 4293 &     K2 \\
57999.0887 & 4294 &     K2 \\
58001.3563 & 4295 &     K2 \\
58003.6239 & 4296 &     K2 \\
58005.8911 & 4297 &     K2 \\
58008.1591 & 4298 &     K2 \\
58010.4253 & 4299 &     K2 \\
58012.6938 & 4300 &     K2 \\
58014.9606 & 4301 &     K2 \\
58017.2283 & 4302 &     K2 \\
58019.4950 & 4303 &     K2 \\
58021.7631 & 4304 &     K2 \\
58024.0301 & 4305 &     K2 \\
58026.2977 & 4306 &     K2 \\
58028.5653 & 4307 &     K2 \\
58030.8330 & 4308 &     K2 \\
58033.1004 & 4309 &     K2 \\
58035.3673 & 4310 &     K2 \\
58037.6349 & 4311 &     K2 \\
58039.9024 & 4312 &     K2 \\
58042.1704 & 4313 &     K2 \\
58044.4379 & 4314 &     K2 \\
58046.7052 & 4315 &     K2 \\
58048.9726 & 4316 &     K2 \\
58051.2393 & 4317 &     K2 \\
58053.5077 & 4318 &     K2 \\
58055.7745 & 4319 &     K2 \\
58058.0413 & 4320 &     K2 \\
58060.3091 & 4321 &     K2 \\
58062.5767 & 4322 &     K2 \\
58064.8452 & 4323 &     K2 \\
58067.1116 & 4324 &     K2 \\
58069.3788 & 4325 &     K2 \\
58071.6470 & 4326 &     K2 \\
58073.9141 & 4327 &     K2 \\
58076.1819 & 4328 &     K2 \\
\enddata
\end{deluxetable*}

\section{Radial Velocity Measurements} \label{sec:RV}
A total of 19 spectra were obtained of the IT Lib system. Sixteen spectra were obtained with the CTIO 1.5 m telescope and CHIRON spectrograph in fiber mode for exposure times of 900 seconds each \citep{2013PASP..125.1336T,Paredes_2021} during 2019 April and May. Three spectra were obtained from the APO 3.5 m telescope and ARCES spectrograph \citep{2003SPIE.4841.1145W} for 1570 seconds each during 2020 June and August. The CTIO spectra covered a wavelength range from 4505 \AA\ to 8895 \AA\ with a resolving power of $R=$ 28,000, and the APO spectra covered a wavelength range from 4015 \AA\ to 7140 \AA\ with a resolving power of $R=$ 30,000. The CTIO CHIRON spectra were reduced following the procedures outlined by \cite{2013PASP..125.1336T} and \cite{2021AJ....162..176P}. The APO data were reduced using the method outlined in the ARC Echelle Spectrograph (ARCES) Data Reduction Cookbook by Karen Kinemuchi. This cookbook describes the steps to use IRAF\footnote{IRAF was distributed by the National Optical Astronomy Observatory, which was managed by the Association of Universities for Research in Astronomy (AURA), under a cooperative agreement with the National Science Foundation.} for cosmic ray removal, bias and flat field subtraction, aperture extraction, and wavelength calibration. The blaze function variation in each echelle order was then removed using a technique from \cite{2015MNRAS.451.4150K} that fits a polynomial to the blaze function for orders with no major absorption, and uses that fit to remove the blaze function from orders with major absorption features such as the H Balmer lines. 

The CHIRON and ARCES spectra were each flux rectified to a unit
continuum and placed into matrices on a uniform heliocentric $\log \lambda$
grid for each observation and echelle order.  Radial velocities were
measured using the two-dimensional cross-correlation code TODCOR
developed by \cite{1994ApJ...420..806Z}, as implemented in \cite{2019AJ....157..140L}.  This code performs a cross-correlation
of the observed and model spectra over a grid of trial velocity shifts
for the primary and secondary stars.  The model template spectra
were formed using theoretical spectra from the TLUSTY BSTAR2006 grid
for the hotter primary \citep{2007ApJS..169...83L} and from the ATLAS BLUERED
spectral library for the cooler secondary \citep{2008AA...485..823B}.
These are all solar abundance model spectra.

The templates were formed by linear interpolation in $(T_{\rm eff}, \log g)$
grids and then were broadened to account for rotational and instrumental
line broadening.   The stellar parameters were initially taken from
\cite{2003PASP..115...49M} and then revised using our own estimates (Tables \ref{tab:ELCVal} and \ref{tab:SpecVal})
on a second iteration.  The TODCOR code creates cross-correlation functions
of the observed and model templates to find the best-fit Doppler shifts
for the primary and secondary, and then the results from all the echelle
orders are combined to determine the final radial velocity and its uncertainty
\citep{2003MNRAS.342.1291Z}.  Our results are given in Table \ref{tab:Phys} that lists the mid-time
of observation, orbital phase (zero at primary eclipse; Section \ref{sec:Per}), radial
velocity and the observed minus calculated residual $O-C$ for both the
primary and secondary, and the source of the measurement.  TODCOR also calculates the flux ratio between the secondary and primary,
$F_2/F_1$, for each echelle order in each observation.  This is measured
from the relative cross correlation peak strengths that are in turn based
upon the adopted spectral template models for each star.  The derived
flux ratio is most sensitive to the deepest spectral lines in each
star's spectrum, and for this purpose we selected an echelle order
in the CHIRON spectra that records the strong H$\beta$ absorption line and
the mainly featureless redward continuum (full range of 4855 to 4909 \AA ).
The monochromatic flux ratio derived from this order
(for a central wavelength of 4882 \AA ) is $F_2/F_1 = 0.38 \pm 0.04$,
where the uncertainty is the standard deviation between observations.
This result is based upon model templates for the effective temperatures
given in Table 5 below.  However, the appearance of the model H$\beta$ line
is sensitive to the adopted temperature, and the line becomes wider
(greater Stark broadening) at lower temperatures.  Thus, if we were
to adopt a lower temperature model for the secondary, the H$\beta$ wings
of the secondary would overlap more with the primary's H$\beta$ feature
in the composite spectrum, so the code would assign less flux to the
primary to account for the changed appearance of the composite spectrum.
We performed tests for this difference by running TODCOR with secondary
spectrum models that cover the $\pm 1 \sigma$ range of effective temperature.
The most significant change occurred by setting the secondary temperature
to the lower boundary value, and the resulting monochromatic flux ratio was
$F_2/F_1 = 0.45 \pm 0.17$.  Thus, the uncertainty quoted previously for
the derived flux ratio is an underestimate that does not account for
the systematic error associated with the adopted spectral models.

The TODCOR code
was successful in determining both component velocities except in the
CHIRON spectrum obtained on BJD 2,458,606.7840 when the components were
too blended for separate measurement. Table \ref{tab:Phys} also includes five measurements
from \cite{2003PASP..115...49M} for cases where both components were measured and
from a single blue archival spectrum from VLT UVES (P.I.\ McEvoy) that
was measured in the same way.

\begin{figure}[h!]
    \begin{center}
        \includegraphics[scale=0.6]{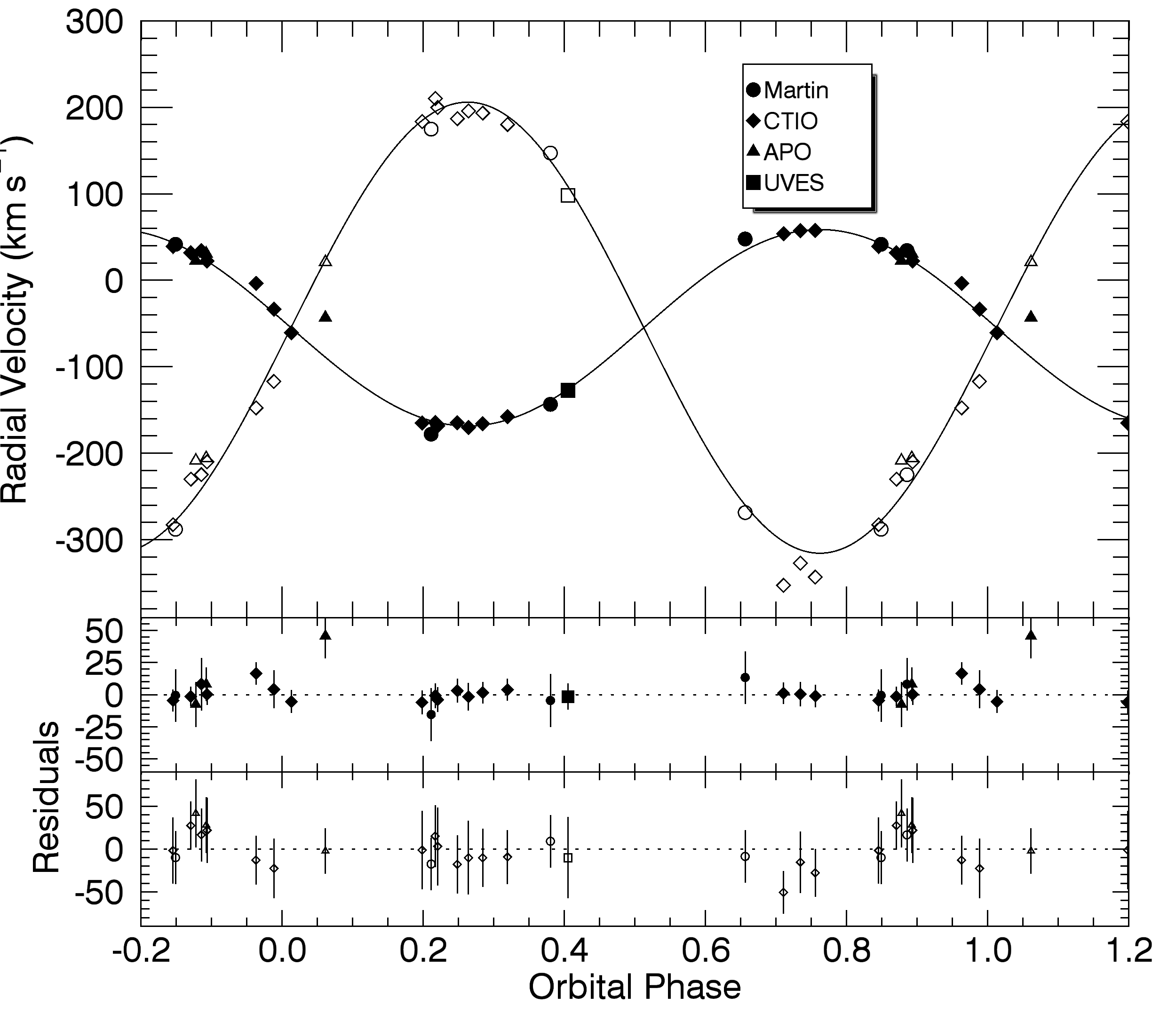}
    \caption{Radial velocity curve for IT Lib. The primary data are shown in black and the secondary data in white. }
    \label{fig:RVcurve}
    \end{center}
\end{figure}

We use our radial velocity measurements along with a list of the orbital phases for the velocities to calculate the orbital parameters for the system, assuming the period ($P$) and epoch ($T_0$) determined earlier. We used the IDL code {\it rvfit.pro} that uses adaptive simulated annealing to find the best Keplerian motion that fits the data \citep{2015ascl.soft05020I} and provides estimates of the uncertainty in the parameters through its Monte Carlo Markov Chain (MCMC) feature. The radial velocity curve is shown in Figure \ref{fig:RVcurve}. There are two CTIO measurements near orbital phase 0 that probably show the Rossiter-McLaughlin Effect. This effect is a slight redshift offset of radial velocity measurements during the
onset of eclipses and a corresponding blueshift offset during egress that is caused by selective obscuration of the background rotating star. Fits excluding these two measurements gave the same orbital solution within uncertainties. The orbital parameter results are presented in Table \ref{tab:paramfit}. These results include the systemic velocity ($\gamma$), the velocity semi-amplitudes ($K_1$, $K_2$), the r.m.s. of the residuals from the fit ($\sigma$), the projected semimajor axis ($a \; \sin i$), and projected masses ($M_1 \sin^{3}i$, $M_2 \sin^{3}i$). The eccentricity was consistent with 0, indicating a circular orbit. Our results are in general agreement with those from \cite{2003PASP..115...49M} (see Table \ref{tab:paramfit}, column 3), although we find a slightly smaller semiamplitude for the primary.
\begin{deluxetable*}{ccccccc}
\tablenum{2}
\tablecaption{IT Lib Radial Velocity Measurements \label{tab:Phys}}
\tablewidth{0pt}
\tablehead{
\colhead{} & \colhead{} & \twocolhead{Primary} & \twocolhead{Secondary} & \colhead{} \\
\colhead{Time} & \colhead{Orbital Phase} & \colhead{$v_{r}$} & \colhead{$O-C$} & \colhead{$v_{r}$} & \colhead{$O-C$} & \colhead{Source} \\
\colhead{(BJD -- 2400000)} & \colhead{} & \colhead{(km~s$^{-1}$)} & \colhead{(km~s$^{-1}$)} & \colhead{(km~s$^{-1}$)} & \colhead{(km~s$^{-1}$)} & \colhead{}
}
\decimalcolnumbers
\startdata
51268.950 & 0.8396 &    \phs $\phn42 \pm   20$& $\phn-16$ &             $-288  \pm  30$ & $-18$ & Martin (2003) \\
51947.000 & 0.8772 &     \phs $\phn34 \pm   20$& $\phn-4$ &             $-225 \pm   30$ & \phs$\phn8$ & Martin (2003) \\
51950.006 & 0.2028 &    $-178 \pm   20$ & \phs$13$ &           \phs$175  \pm 30$ & $\phn-7$ & Martin (2003) \\
51951.015 & 0.6479 &    \phs $\phn48 \pm  20$& \phs$\phn0$ &			  $-269 \pm 30$ & $-11$ & Martin (2003) \\
52419.751 & 0.3725 &    $-143 \pm  20$& \phs$\phn9$ & 				\phs$147  \pm  30$ & \phs$15$ & Martin (2003) \\
56734.761 & 0.4051 &    $-127 \pm 10$ & $\phn-1$ & \phs$\phn 98 \pm 47$ & $-11$ &VLT UVES \\
58593.7484 & 0.2648 &   $-170 \pm  10$& $\phn-2$ &            \phs$196 \pm  42$ & $\phn-11$ & CTIO 1.5 m CHIRON\\
58593.7942 & 0.2850 &   $-166 \pm   \phn8$& \phs$\phn1$ &            \phs$193\pm   33$ & $-11$ & CTIO 1.5 m CHIRON\\
58593.8739 & 0.3202 &   $-158 \pm   \phn8$& \phs$\phn4$ &            \phs$180  \pm 31$ & $-11$ & CTIO 1.5 m CHIRON\\
58594.7608 & 0.7113 &  \phs   $\phn54 \pm   \phn8$& \phs$\phn1$ &           $-353 \pm  24$ & $-50$ & CTIO 1.5 m CHIRON\\
58594.8150 & 0.7352 &  \phs   $\phn57 \pm  9$& \phs$\phn0$ &           $-327  \pm 35$ & $-15$ & CTIO 1.5 m CHIRON\\
58594.8629 & 0.7564 &  \phs   $\phn58 \pm   \phn8$& $\phn-1$ &           $-343 \pm  27$ & $-28$ & CTIO 1.5 m CHIRON\\
58606.6702 & 0.9637 &     $ \phn -4 \pm   \phn8$& \phs$17$ &           $-148   \pm 28$ & $-14$ & CTIO 1.5 m CHIRON\\
58606.7278 & 0.9890 &    $\phn -34 \pm  14$& \phs$\phn4$ &           $-117\pm   34$ & $-24$ & CTIO 1.5 m CHIRON\\
58606.7840 & 0.0139 &   $\phn -61 \pm  8$& $\phn-5$ &           \nodata & \nodata & CTIO 1.5 m CHIRON\\
58608.6711 & 0.8461 &  \phs   $\phn39 \pm   \phn8$& $\phn-4$ &           $-283 \pm  38$ & $\phn-3$ & CTIO 1.5 m CHIRON\\
58608.7277 & 0.8714 &  \phs   $\phn32 \pm   \phn7$& $\phn-1$ &           $-230  \pm 27$ & \phs$26$ & CTIO 1.5 m CHIRON\\
58608.7803 & 0.8943 &  \phs   $\phn22 \pm   \phn8$& \phs$\phn1$ &           $-210 \pm  37$ & \phs$20$ & CTIO 1.5 m CHIRON\\
58611.7891 & 0.2212 &   $-168 \pm  9$& $\phn-4$ &            \phs$200 \pm  45$ & \phs$\phn3$ & CTIO 1.5 m CHIRON\\
58611.8522 & 0.2491 &   $-165 \pm  9$& \phs$\phn3$ &            \phs$187\pm  33$ & $-18$ & CTIO 1.5 m CHIRON\\
58620.8087 & 0.1991 &   $-165 \pm   9$& $\phn-6$ &            \phs$184  \pm 45$ & $\phn-1$ & CTIO 1.5 m CHIRON\\
58620.8511 & 0.2178 &   $-165 \pm  9$& $\phn-1$ &            \phs$210\pm   35$ & \phs$15$ & CTIO 1.5 m CHIRON\\
59007.8151 & 0.8789 &  \phs   $\phn22 \pm  17$& $\phn-7$ &           $-209  \pm 39$ & \phs$40$ & APO 3.5 m ARCES\\
59073.6044 & 0.8937 &  \phs   $\phn31 \pm  13$& \phs$\phn9$ &           $-206\pm   32$ & \phs$26$ & APO 3.5 m ARCES\\
59087.5921 & 0.0627 &    $\phn -43 \pm  17$& \phs$45$ &             \phs$\phn21 \pm  26$ & $\phn-2$ & APO 3.5 m ARCES\\
\enddata
\end{deluxetable*}

\begin{deluxetable*}{ccc}
\tablenum{3}
\tablecaption{IT Lib Orbital Parameters ($e = 0$) \label{tab:paramfit}}
\tablewidth{0pt}
\tablehead{
\colhead{Parameter} & \colhead{This Paper} & \colhead{Martin (2003)}
}
\decimalcolnumbers
\startdata
$P$ (days) & $2.2674456 \pm 0.0000003$ &  $2.2674600$\\
$T_0$ (days) & $2448262.6772 \pm 0.0014$ BJD & $2448500.76777$ JD\\
$\gamma \; $(km~s$^{-1})$ & $-54.7  \pm 2.4$ & $-50.9 \pm 7.0$\\
$K_{1} \; $(km~s$^{-1})$ & $113.5 \pm 2.7$ & $122.7^{a}$ \\
$K_{2} \; $(km~s$^{-1})$ & $261.0 \pm 9.4$ & $262.6^{a}$ \\
$\frac{M_{2}}{M_{1}}$ & $0.435 \pm 0.019$ & $0.467 \pm 0.044$\\
$a \; \sin i \; (R_{\odot})$ & $16.79 \pm 0.44$ & $17.22^{a} $\\
$M_{1} \sin^{3}i \; (M_{\odot})$ & $8.61 \pm 0.55$ & $9.15^{a} $\\
$M_{2} \sin^{3}i \; (M_{\odot})$ & $3.74 \pm 0.21$ & $4.28^{a}$\\
$\sigma_1 \; $(km~s$^{-1})$ & 11 & \nodata \\
$\sigma_2 \; $(km~s$^{-1})$ & 20 & \nodata \\
\enddata
\tablecomments{Period and epoch were fixed in the solution.}
\tablenotetext{a}{These values are derived from the results of Martin (2003).}
\end{deluxetable*}

\section{Light Curve Modeling} \label{sec:ELC}
The fit of the radial velocity curve provides many of the orbital parameters, but the inclination is needed to determine the masses from the radial velocity results. We can find the inclination and other parameters by fitting the Kepler K2 light curve with the light curve models generated by the Eclipsing Light Curve (ELC) code \citep{2000AA...364..265O}. The ELC models require parameters determined from the spectroscopic orbit including the period, epoch, velocity semiamplitudes, and projected separation. Setting these spectroscopically derived values allows ELC to just vary the values of only four parameters to fit the light curve: the inclination ($i$), radii ($R_1 , R_2$), and temperature ratio ($T_2 / T_1$) of the system. We use the genetic algorithm optimization method to solve for these, and the results are given in Table \ref{tab:ELCVal}. The genetic algorithm method takes inspiration from natural selection by creating multiple lineages where each lineage is composed of different combinations of values for the parameters in question. ELC will create a model based on these values, and compare the fit of each lineage's model to the data. The worst fitting lineages are killed (deleted and not used thereafter), but the best fitting lineages go onto the next generation with mutations. The killed lineages are replaced by lineages with new random parameters or mutations of successful lineages. In this way, the worst fitting sections of the parameter space are avoided, but the more successful regions are explored. In addition, the random nature of the mutations also helps avoid being caught in local minima of the parameter space and to find the true global minimum. 

ELC uses the selected parameters to create a model of the Roche-distorted
shapes and surface temperature distributions and to derive the total monochromatic
flux of the system as a function of viewing angle (dependent on inclination
and orbital phase).  We used the plane-parallel, LTE atmosphere models from
the ATLAS code by Robert Kurucz to create a table of specific intensities
for the Kepler-band mean wavelength as functions of local temperature, gravity,
and viewing angle relative to the surface normal.  The ELC code calculates
the integrated flux from all non-occulted surface elements to create a
monochromatic flux estimate in orbital phase steps of 0.11 deg (equivalent to about 1 minute of time in the binary orbit).  Then the resulting model light
curve is smoothed with a boxcar function to account for the 29.4 minute
integration time of the Kepler long cadence photometric observations.
The Kepler photometric fluxes were averaged into bins of 0.01 of orbital phase,
and the ELC model curve was re-normalized to the rebinned light curve
before calculating the $\chi^2$ goodness-of-fit.  The derived inclination
depends mainly on the eclipse depths, and the temperature ratio follows
from the relative depths of the primary and secondary eclipses.
The stellar radii depend on the eclipse duration and on the degree of
tidal distortion as observed in the flux variation outside of the eclipses.
We used ELC to explore the full range of possible radii, and the best fits
were unambiguously obtained in models where the secondary fills its Roche
lobe and its tidal distortion creates the out-of-eclipse flux variation. This large radius is also supported by our spectroscopy result from TODCOR. The monochromatic flux ratio that best matches the spectral line depths reveals that the secondary is relatively bright. From ELC, we found that the monochromatic flux ratio for our best fit model is $F_2 / F_1 = 0.38$ at $6650$ \AA\ (the central wavelength of the Kepler bandpass), and from spectroscopy we found the flux ratio to be $F_2 / F_1 = 0.38 \pm 0.04$ at $4882$ \AA . Both the primary and secondary stars are hot enough that the $4882$ \AA\ and $6650$ \AA\ wavelengths are in the Rayleigh-Jeans tail of the flux distribution, so the flux ratio should not change much between them. The fact that the two different observables (light curve modelling and spectroscopy) lead to the same result is strong evidence for a relatively large, Roche-filling secondary.

The next step is to determine the parameter uncertainties. We know the period and epoch extremely well from our long timeline of eclipse timings, so the uncertainties from those are insignificant for the ELC results. The errors in the fluxes are much smaller than the fluxes themselves, with typical flux values of 2.5 million electrons per second and errors of about 45 electrons per second. The determination of uncertainties was done by fixing all of the parameters to their best fit value from the optimization procedure and slightly adjusting one value at a time to be larger and smaller until the $\chi^{2}$ statistic increases by one. If the uncertainty decreases instead, the parameter is adjusted until the uncertainty starts increasing, and the value that gives the lowest uncertainty is chosen as the new best fit value.  Once this was done for all four parameters, the process was repeated for the first parameters to make sure the results did not change when the last parameters were adjusted to better fitting values. The $\chi^{2}$ statistic describing the fit of the model ELC light curve to the Kepler light curve was large, so it was re-normalized to be equal to the number of observed datapoints in regions most sensitive to changes in the parameters. 

ELC finds the fractional radii directly, but converting those into stellar radii requires knowing the physical size scale of the binary. The physical size scale depends directly on the semiamplitudes and resulting semimajor axis from the radial velocity fitting results. So, the radii have additional uncertainty due to the uncertainty of the scale of the system ($a \sin i$) from the spectroscopic orbit. In fact, this uncertainty dominates the error budget for our radii. The best solution for the secondary radius was the Roche-filling configuration, so our method of uncertainty estimation could only be applied in the direction of the smaller radius. Consequently, we simply assigned the fitting error for the primary radius to the case of the secondary. This decision is insignificant for the error budget which is dominated by the uncertainty in the spectroscopically determined $a \sin i$. Likewise, the uncertainty for the inclination was used with the uncertainty of the mass products from the spectroscopic orbit to calculate the propagated uncertainty in the masses. 

\begin{figure}[h!]
    \begin{center}
        \includegraphics[scale=0.9]{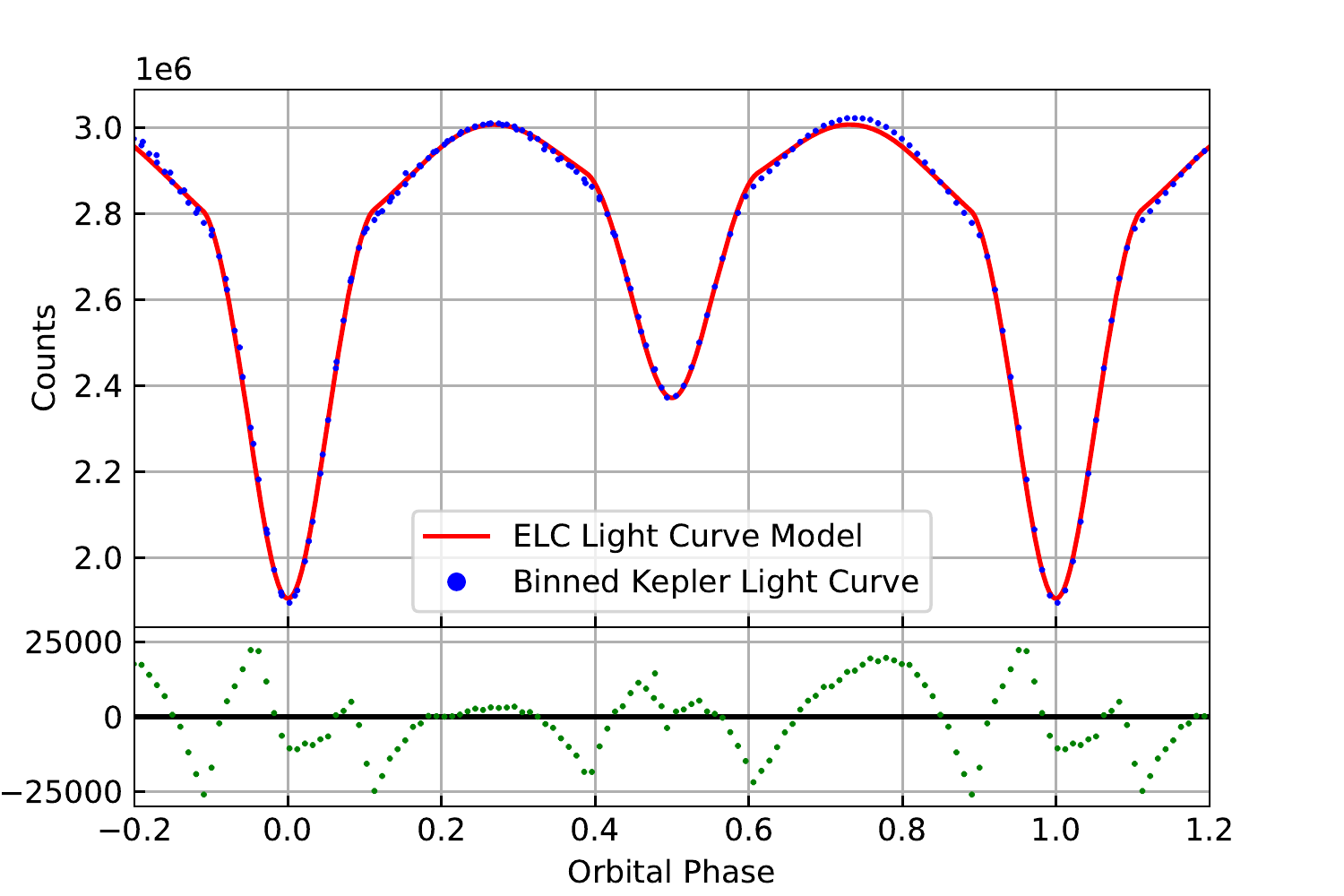}
    \caption{Comparison of the Kepler K2 light curve and the model ELC light curve for the best fit parameters. The Kepler K2 light curve has been binned in phase to ease the comparison between the model and data. Phase zero is the time of primary eclipse. The flux is in units of electrons per second. The bottom section shows the residuals from the fit.}
    \label{fig:LCPlot}
    \end{center}
\end{figure}

Figure \ref{fig:LCPlot} compares the Kepler K2 light curve to the generated model light curve from ELC. There are slight differences between the model and the data for orbital phases outside the eclipse regions, and there is an asymmetry that appears with the second maximum brighter than the first maximum. ELC assumes axial symmetry, so the model cannot account for the difference in the maxima. The reason for the brighter second maximum is unknown, but it may arise from unseen mass transfer heating up the surface of the primary at the point of gas stream impact and causing it to be brighter. The inclination, radii, and temperature ratio found from ELC, along with the resulting masses are shown in Table \ref{tab:ELCVal}. The reported radii are equivalent to those for spheres of volume equal to the Roche distorted volumes of the stars. Table \ref{tab:ELCVal} also lists the corresponding results from \cite{2003PASP..115...49M}. The sum of the radii is similar to that found by \cite{2003PASP..115...49M}, but we find a larger (Roche filling) secondary compared to his result. Model light curves with a smaller secondary radius (and larger primary radius) fail to match the observed light curve in the non-eclipse phases.

\begin{deluxetable*}{ccc}
\tablenum{4}
\tablecaption{IT Lib Light Curve Parameters \label{tab:ELCVal}}
\tablewidth{0pt}
\tablehead{
\colhead{Parameter} & \colhead{This Paper} & \colhead{Martin (2003)}
}
\decimalcolnumbers
\startdata
$R_{1} \; (R_{\odot})$ & $6.06 \pm 0.16$ & $6.87 $\\
$R_{2} \; (R_{\odot})$ & $5.38 \pm 0.14$ & $4.77 $\\
$\frac{T_{2}}{T_{1}}$ & $0.6458 \pm 0.0007$ & 0.62 \\
$i \; $(degrees) & $74.61 \pm 0.02$ & $78.1 \pm 3.5$\\
\tableline
$M_{1} \; (M_{\odot})$ & $9.60 \pm 0.56$ & $9.77 \pm 0.65$\\
$M_{2} \; (M_{\odot})$ & $4.18 \pm 0.21$ & $4.57 \pm 0.30$\\
$a\; (R_{\odot})$ & $17.41 \pm 0.46$ & $17.6$\\
\enddata
\tablecomments{The first four parameters were the fitting parameters, and the last three parameters (the masses and semimajor axis) were derived from the inclination and spectroscopic orbit.}
\end{deluxetable*}

\newpage
\section{Spectral Analysis} \label{sec:Spec}
The next step is to analyze the spectra of the system to derive stellar atmospheric parameters. The primary and secondary star spectra are reconstructed using Doppler Tomography as described in \cite{1992ApJ...385..708B}. It uses model templates for each star, the radial velocities associated with each star for each spectrum, and a monochromatic flux ratio for the stars to create the best estimate for the individual spectra of each star. The models are from the same TLUSTY and BLUERED grids described in Section \ref{sec:RV}, computed for solar metallicity and a microturbulence value of 2 km~s$^{-1}$ (suitable for main sequence stars). We primarily used the CHIRON spectra for this analysis and the radial velocities used are given in Table \ref{tab:Phys}. We then made fits of the reconstructed spectra to determine their effective temperatures and projected rotational velocities $V \sin i$, assuming gravities of $\log g = 3.8$ for the primary and $\log g =3.6$ for the secondary, as derived from the ELC results given in Table \ref{tab:ELCVal}. We estimate the temperatures $T_{\rm eff}$ and projected rotational velocities by constructing a grid of models for both parameters. The code described by \citet{2019AJ....157..140L} combines the test estimates of the model primary and secondary spectra and uses a cross-correlation function to evaluate how well they match the reconstructed spectra. The code determines the best fitting primary and secondary temperatures with uncertainties and the best fitting rotational velocities with uncertainties. The synchronous projected rotational velocity ($V_s \sin i$) is also calculated (using the period, inclination and radius results from ELC) and presented for comparison.  These are listed in Table \ref{tab:SpecVal}. It is encouraging to see that the spectroscopic temperature ratio $T_2 / T_1 = 0.58 \pm 0.11$ agrees within errors with the light curve determined temperature ratio $T_2 / T_1 = 0.6458 \pm 0.0007$, and that our temperatures are in the range expected for the spectral type of B2/3 assigned by \citet{1988mcts.book.....H} for the brighter primary star.

\begin{figure}[h!]
    \begin{center}
        \includegraphics[scale=0.6, clip, trim=0cm 7cm 0cm 8.5cm]{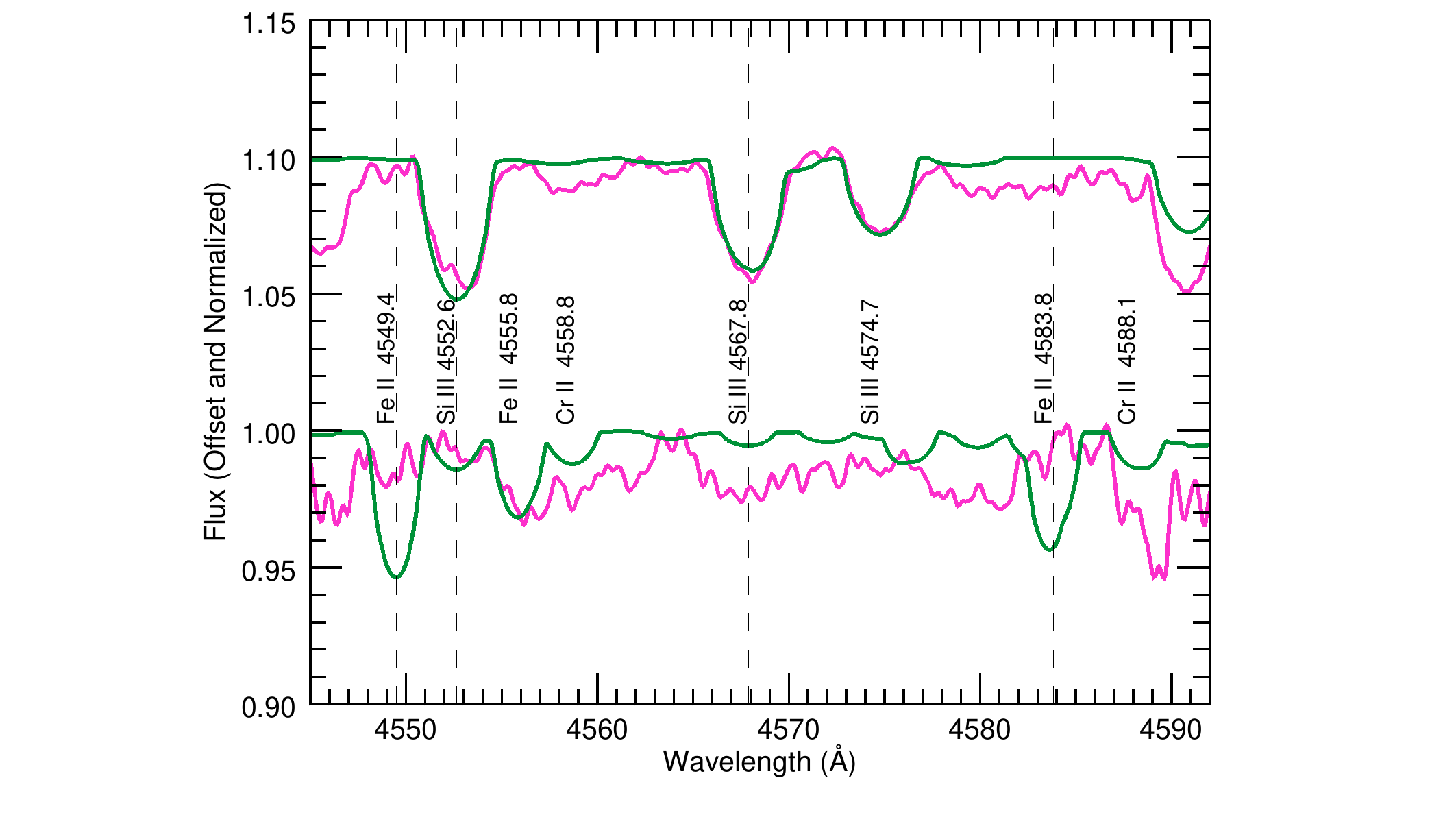}
    \caption{Reconstructed and model spectra for the wavelength range of $4542$ to $4592$ \AA. The reconstructed spectra are in pink and the model spectra are in green with the primary component being offset to a larger normalized flux. The order showcases the \ion{Si}{3} $\lambda\lambda 4552, 4567, 4574$ absorption features.}
    \label{fig:s1}
    \end{center}
\end{figure}

\begin{figure}[h!]
    \begin{center}
        \includegraphics[scale=0.6, clip, trim=0cm 7cm 0cm 8.5cm]{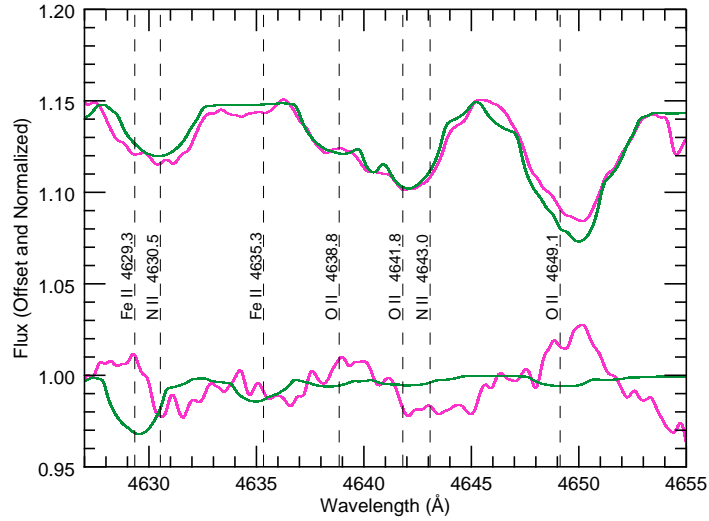}
    \caption{Reconstructed and model spectra for the wavelength range of $4627$ to $4655$ \AA. The reconstructed spectra are in pink and the model spectra are in green with the primary component being offset to a larger normalized flux. The order showcases the \ion{N}{2} $\lambda\lambda 4630, 4643$ and \ion{O}{2} $\lambda\lambda 4638, 4641, 4649$ (plus other weaker blends).}
    \label{fig:s3}
    \end{center}
\end{figure}

\begin{figure}[h!]
    \begin{center}
        \includegraphics[scale=0.6, clip, trim=0cm 7cm 0cm 8.5cm]{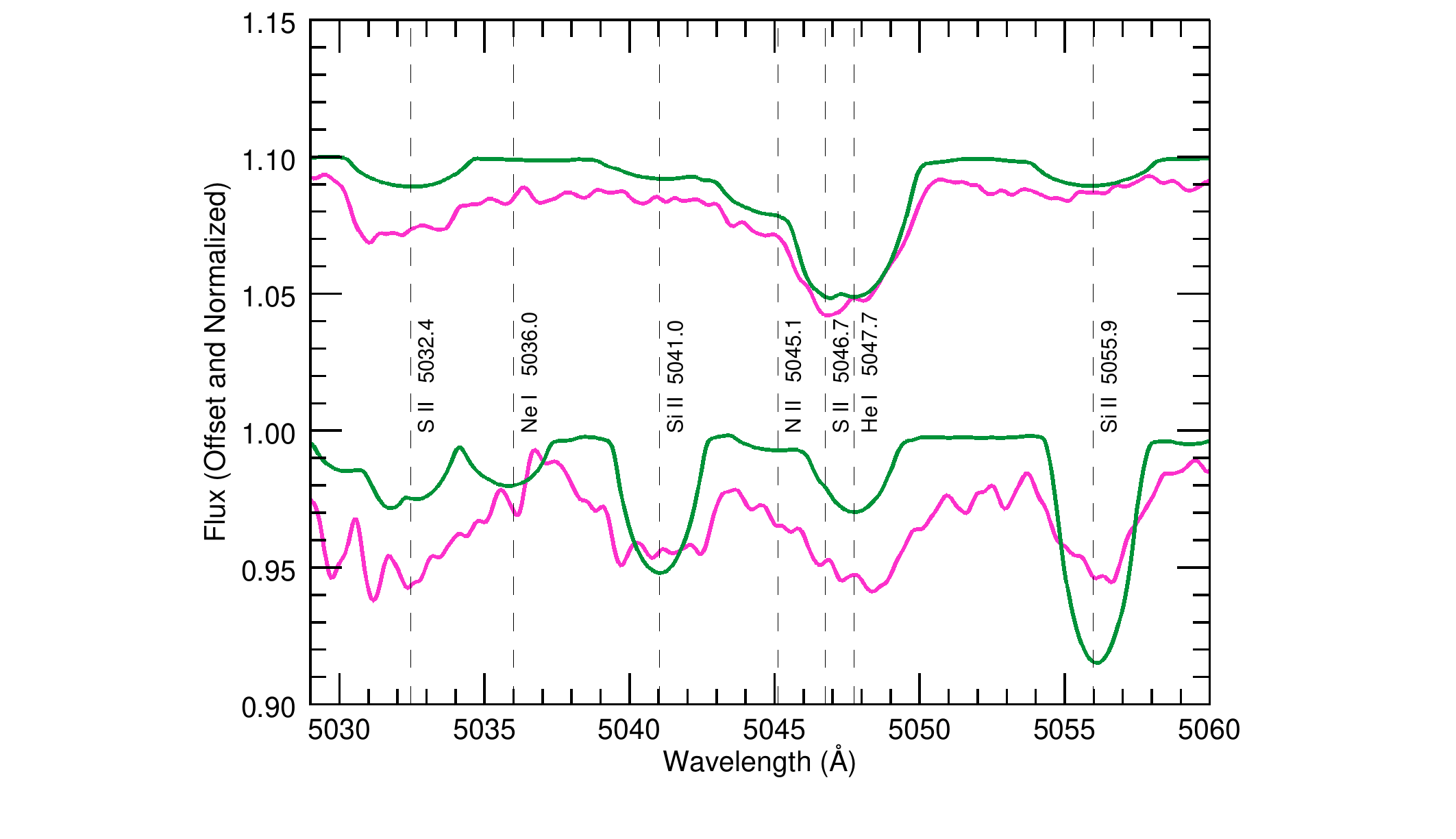}
    \caption{Reconstructed and model spectra for the wavelength range of $5029$ to $5060$ \AA. The reconstructed spectra are in pink and the model spectra are in green with the primary component being offset to a larger normalized flux. The order showcases the \ion{He}{1} $\lambda 5047$, \ion{N}{2} $\lambda 5045$, \ion{S}{2} $\lambda 5032$, and \ion{Si}{2} $\lambda\lambda 5041, 5056$ absorption features.}
    \label{fig:s13}
    \end{center}
\end{figure}

\begin{figure}[h!]
    \begin{center}
        \includegraphics[angle=90,scale=0.45, clip, trim=0cm 0cm 0cm 0cm]{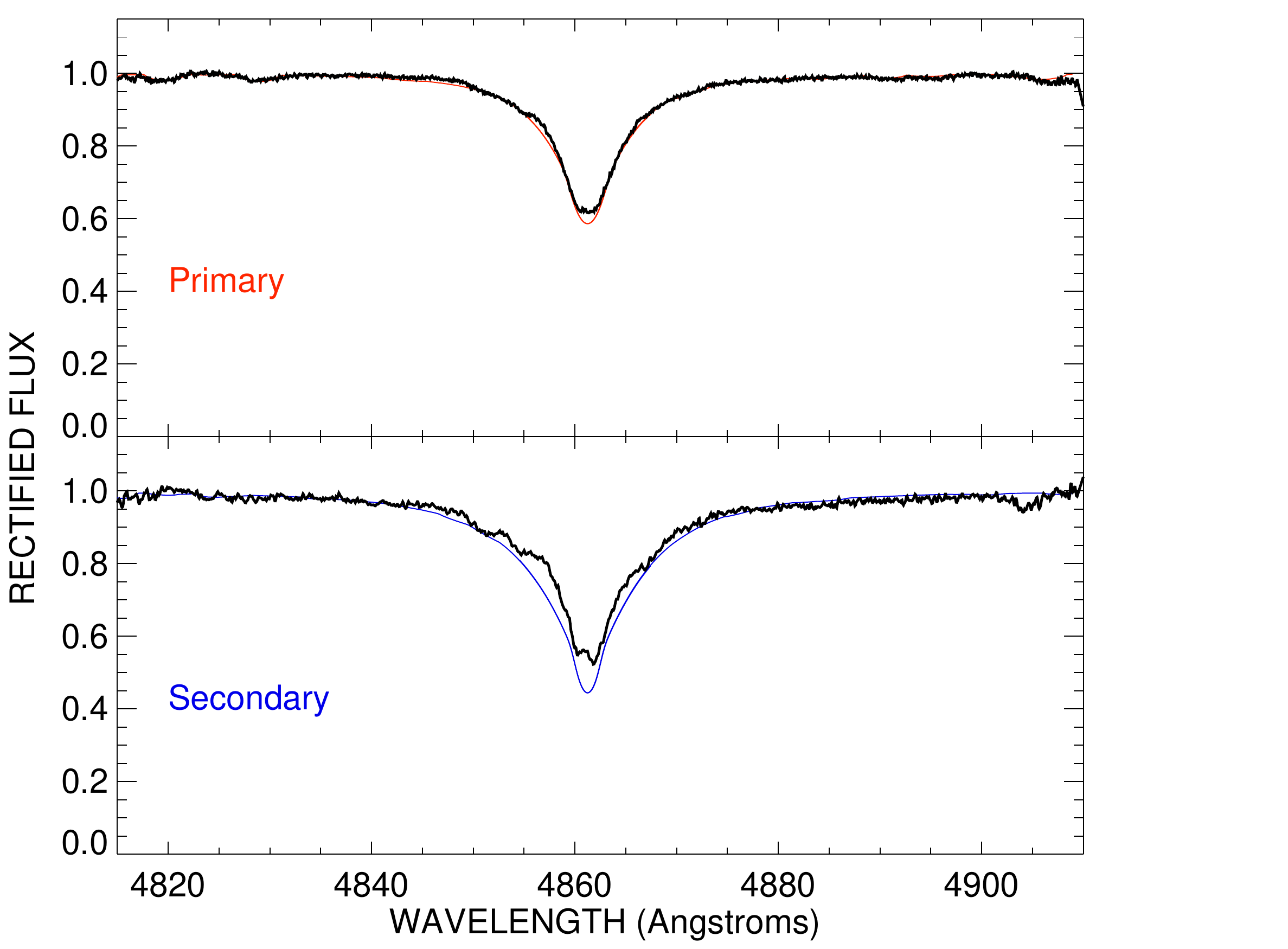}
    \caption{Combined reconstructed and model spectra for the wavelength range of $4815$ to $4910$ \AA\, from two adjacent echelle orders. The reconstructed spectra are in black and the model spectra are in red for the primary and blue for the secondary. This plot showcases the H$\beta$ $\lambda 4861$ absorption feature. The absorption feature for the secondary may show the impact of in-filling due to mass transfer.}
    \label{fig:s9}
    \end{center}
\end{figure}

\begin{figure}[h!]
    \begin{center}
        \includegraphics[scale=0.6, clip, trim=0cm 7cm 0cm 8.5cm]{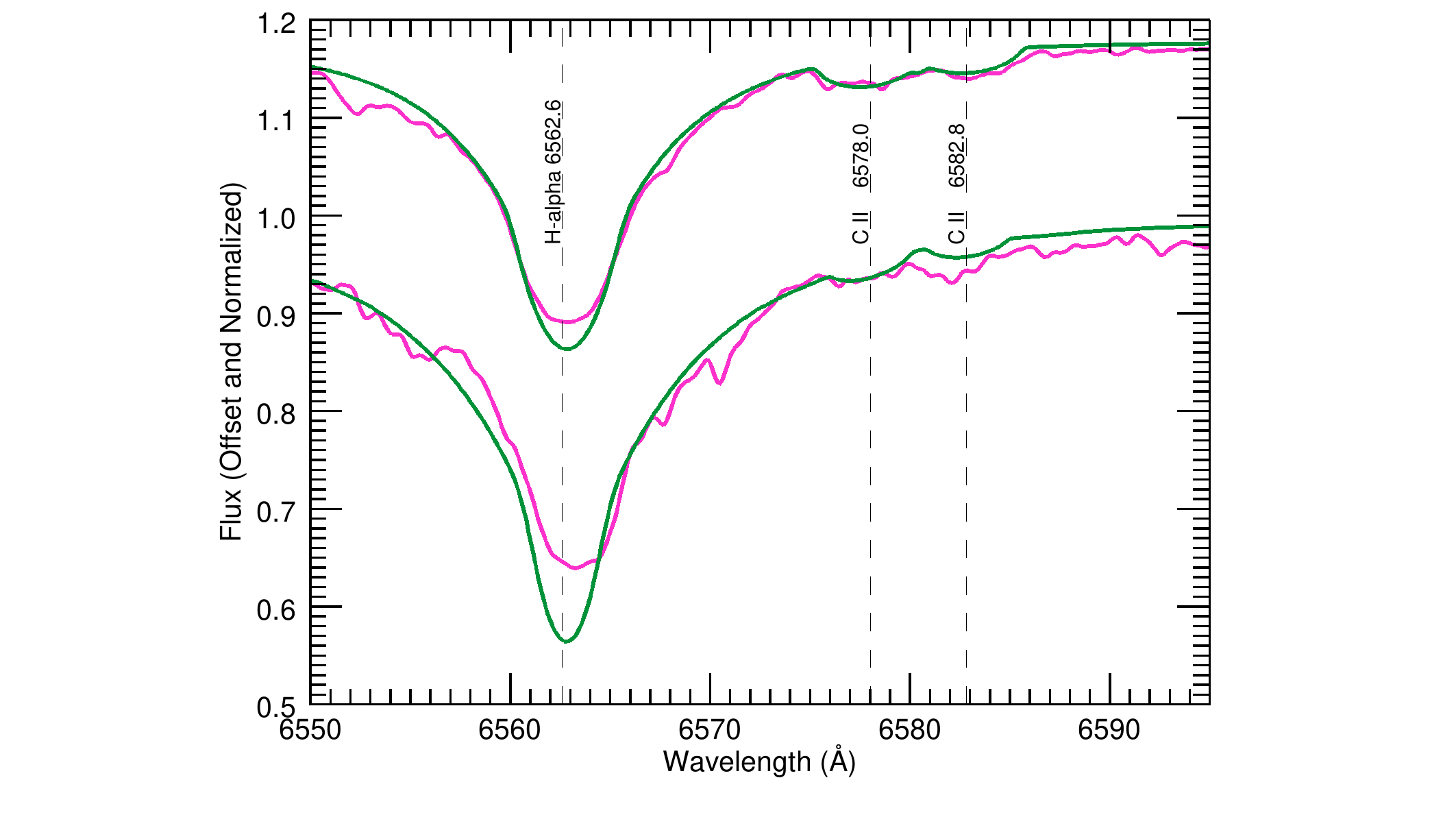}
    \caption{Reconstructed and model spectra for the wavelength range of $6550$ to $6595$ \AA. The reconstructed spectra are in pink and the model spectra are in green with the primary component being offset to a larger normalized flux. The order showcases the H$\alpha$ $\lambda 6563$ and \ion{C}{2} $\lambda\lambda 6578, 6582$ absorption features.}
    \label{fig:s39}
    \end{center}
\end{figure}

All of the reconstructed spectra are available from the authors. A subset of these are given here as they are especially useful for our analysis. These include Figures \ref{fig:s1}, \ref{fig:s3}, \ref{fig:s13}, \ref{fig:s9}, and \ref{fig:s39}. In Figures \ref{fig:s1}, \ref{fig:s3}, \ref{fig:s13}, and \ref{fig:s39}, the reconstructed spectra are in pink and the model spectra are in green, with the primary component offset to a larger normalized flux. In Figure \ref{fig:s9}, the reconstructed spectra are in black with the primary model spectrum in red and the secondary model spectrum in blue. The reconstructed spectra were smoothed and rescaled in intensity and wavelength for ease of inspection in these figures. An analysis of the segments of the spectra can give some insight on the stellar properties. We begin with a discussion of the primary star. Figure \ref{fig:s1}, shows the \ion{Si}{3} $\lambda\lambda 4552, 4567, 4574$ lines which are fit very well by the model for the primary. The \ion{Si}{3} lines are a powerful temperature diagnostic for hot B stars, as they disappear for the hottest B stars (replaced by \ion{Si}{4}) and the cooler B stars (replaced by \ion{Si}{2}). They also should not be affected by non-solar He and CNO abundances that may be altered by mass transfer. Figure \ref{fig:s3} shows both the \ion{N}{2} $\lambda 4631$ and \ion{O}{2} $\lambda 4649$ lines. These are the important elements of the CNO cycle occuring within these massive stars. During the CNO cycle, the amount of nitrogen is enhanced and so it might be expected that the mass gainer star will have an excess of nitrogen due to mass transfer from the donor star. However, the model fit to the primary reconstructed spectra shows nothing out of the ordinary for these species. Thus, our results are consistent with the assumption of solar abundances of O and N for the primary as also found by \citet{2003PASP..115...49M}.

We now move to our discussion of the secondary. We do not see the \ion{Si}{3} $\lambda\lambda 4552, 4567, 4574$ lines in the secondary spectrum, which is consistent with the cooler temperature of the secondary. Figure \ref{fig:s13} shows the lines of \ion{He}{1} $\lambda 5047$ and \ion{Si}{2} $\lambda\lambda 5041, 5056$. The \ion{Si}{2} lines are a fairly good match to the model, and like the \ion{Si}{3} lines, they are a good temperature indicator that is less vulnerable to confusion from mass transfer effects. They become weaker at hotter B star temperatures and at cooler A star temperatures. While these Si lines are well fit, \ion{He}{1} $\lambda 5047$ is significantly stronger in the observed spectrum than in the model one. This enhancement of \ion{He}{1} is also observed in other \ion{He}{1} lines in the secondary spectrum and may indicate an enriched He abundance.

Finally we discuss the Balmer lines. Figure \ref{fig:s9} shows the H$\beta$ line from the flux-weighted average of two adjoining echelle orders. We can see that the secondary's H$\beta$ line is weaker than the model, perhaps suggesting weak emission from mass transfer that tends to fill in the absorption line of the secondary. This same difference is observed in H$\alpha$ shown in Figure \ref{fig:s39}. The H$\alpha$ line is fit for the primary fairly well, but the secondary shows a weaker H$\alpha$ line than the model suggests. This observation, when paired with the He I excess observation, suggests that the model temperature for the secondary is either too cool or that mass transfer is creating weak emission that fills in the hydrogen absorption lines. However, the model temperature for the secondary makes a good fit of the \ion{Si}{2} lines, and tests with hotter temperatures generally led to worse fits.

\begin{deluxetable*}{ccc}
\tablenum{5}
\tablecaption{IT Lib Spectroscopic Temperatures and Rotational Velocities \label{tab:SpecVal}}
\tablewidth{0pt}
\tablehead{
\colhead{Parameter} & \colhead{This Paper} & \colhead{Martin (2003)}
}
\decimalcolnumbers
\startdata
$T_{1} \; $(kK) & $23.8 \pm 1.8$ & $22.0$ \\
$T_{2} \; $(kK) & $13.7 \pm 2.5$ & $13.6 \pm 2.1$\\
$V_{1}\sin\;i \;$(km~s$^{-1}) $ & $125 \pm 17$ & \nodata \\
$V_{2}\sin\;i \;$(km~s$^{-1}) $ & $91 \pm 32$ & \nodata \\
$Vs_{1}\sin\;i \;$(km~s$^{-1}) $ & $139$ & \nodata  \\
$Vs_{2}\sin\;i \;$(km~s$^{-1}) $ & $117$ & \nodata \\
\enddata
\end{deluxetable*}

\section{Spectral Energy Distribution and Distance} \label{sec:SED}
We can calculate the distance to IT Lib by fitting its SED to a model binary star spectrum. The observed fluxes of the SED were collected from various sources in the {\it VizieR} photometry tool written by Anne-Camille Simon and Thomas Boch\footnote{http://vizier.u-strasbg.fr/vizier/sed/doc/}. These include fluxes from magnitude measurements from 0.3 to 11.6 $\mu$m in wavelength in the following bands: 
Str\"{o}mgren-Crawford $uvby\beta$ \citep{Paunzen2015}, 
Johnson $U$ from XHIP \citep{Anderson2012}, 
Johnson $BV$ from APASS \citep{Henden2015},  
PAN-STARRS $izy$ \citep{Chambers2016}, 
2MASS $JHK_S$ (from the compliation by \citealt{Kharchenko2001}), 
and AllWISE \citep{Cutri2014}. 

A low resolution model spectrum was created for the binary system by interpolation in ($T_{\rm eff}, \log g$) among Kurucz ATLAS9 LTE model spectra for both the primary and secondary components. The temperatures were set from our spectra analysis and the surface gravities from ELC (derived from the masses and radii in Table \ref{tab:ELCVal}). These models were computed for solar metallicity and a microturbulence value of 2 km~s$^{-1}$ (suitable for main sequence stars). The flux of the secondary was rescaled to that of the primary assuming a monochromatic flux ratio of $F_2/F_1 = 0.38$ in the continuum region near H$\beta$ $\lambda 4861$, and then the primary and secondary star fluxes were added to obtain a model binary star spectrum. 

\begin{figure}[h!]
    \begin{center}
        \includegraphics[scale=0.45]{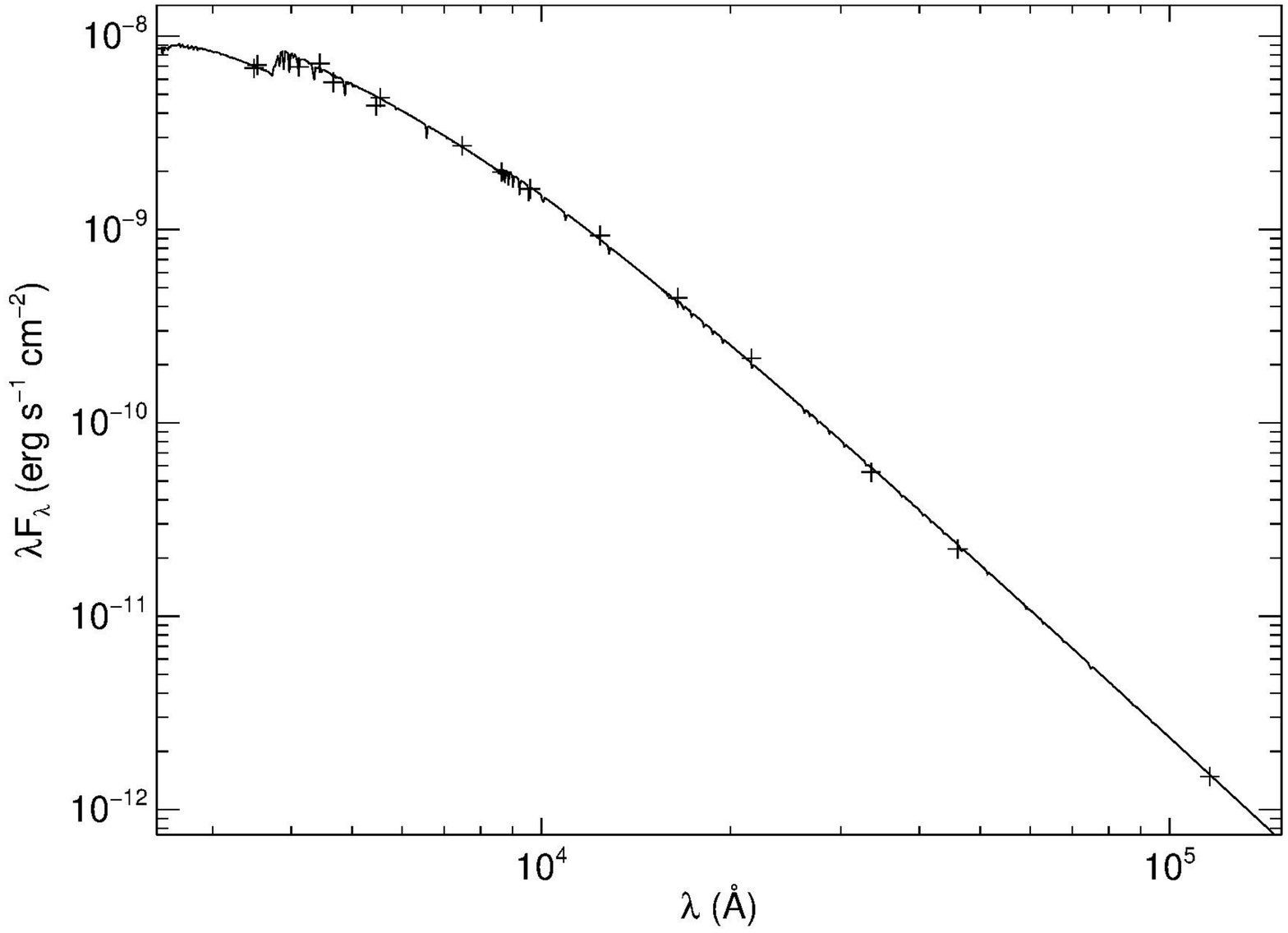}
    \caption{Fit to the observed spectral energy distribution of IT Lib using temperatures derived from the spectral analysis of the component stars.}
    \label{fig:SpecSED}
    \end{center}
\end{figure}

The model was fit to the observed fluxes after applying interstellar extinction for a ratio of total-to-selective extinction of $R_V = 3.1$ using the model extinction curve described by \citet{Fitzpatrick1999}.  A good fit was made with a reddening of $E(B-V) = 0.190 \pm 0.012$ mag and a limb-darkened angular diameter of the primary star of $\theta = 0.0243 \pm 0.0007$ milliarcsec. The observed fluxes and the fit of the spectral energy distribution are shown in Figure \ref{fig:SpecSED}. The distance can be derived from $\theta$ and the average radius of the primary as derived from the ELC light curve solution. 

Finally, the systematic error in the distance was found by using the temperatures at the maximum and minimum values allowed by their uncertainties. In addition to the spectra-derived temperatures, we also have ELC-derived temperatures, from the temperature ratio of ELC and the primary temperature used in ELC. We get the same distance within errors when using either the spectra-derived or ELC-derived temperatures. The estimated distance is $2.32 \pm 0.12$ kpc, which is in good agreement with that from Gaia EDR3, $2.29^{+0.19}_{-0.24}$ kpc \citep{2021AJ....161..147B}. 

\section{Travel Time Determination} \label{sec:TTDet}
Perhaps the most striking result from the analysis of the system by \cite{2003PASP..115...49M} is that the travel time required for IT Lib to reach its current position from the galactic plane exceeds the main sequence lifetime for a star with the mass of the primary. In order to confirm this discrepancy, we consulted MESA evolutionary models \citep{2015ApJS..220...15P} for the component stars in IT Lib to determine their main sequence lifetimes. Specifically, we used MESA-Web\footnote{http://www.astro.wisc.edu/$^\sim$townsend/static.php?ref=mesa-web} \citep{Fields2015}, a web-based interface to the stellar evolution code, to construct evolutionary tracks for both stars with masses of $9.6 M_{\odot}$ and $4.18 M_{\odot}$. We found that the main sequence lifetime is about 21 Myrs for the primary and 130 Myrs for the secondary.

\begin{figure}[h!]
    \begin{center}
        \includegraphics[scale=0.8]{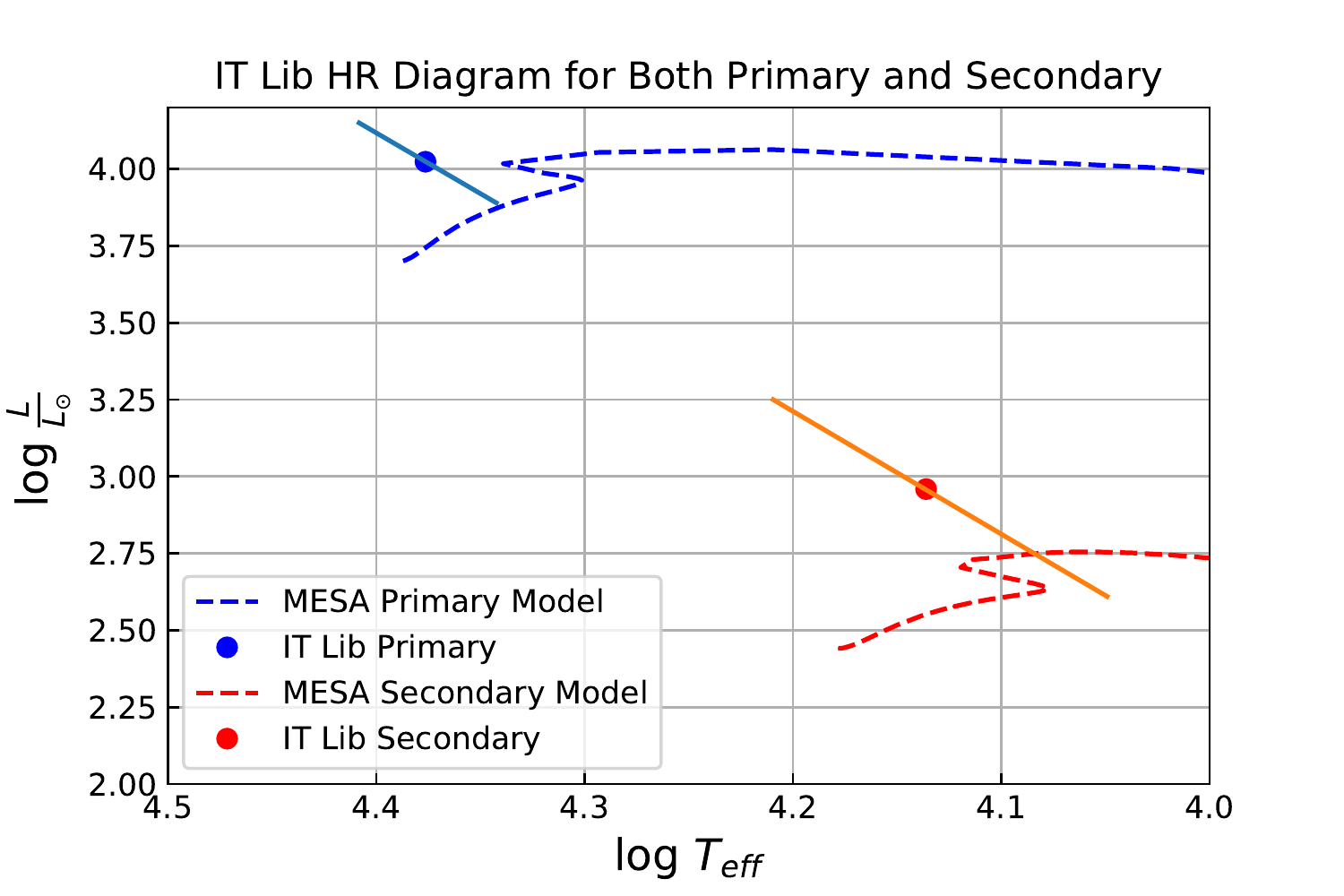}
    \caption{An H-R Diagram for stars of IT Lib's primary and secondary masses, $9.6 M_{\odot}$ and $4.18 M_{\odot}$. Overlaid on each evolutionary track are the stars' actual luminosities and temperatures, along with the uncertainty. Notice that the primary, and in particular the secondary, are above the evolutionary track. This suggests that its evolution was altered through binary mass exchange (Section \ref{sec:Orig}).}
    \label{fig:HR}
    \end{center}
\end{figure}

Figure \ref{fig:HR} is a Hertzsprung-Russell Diagram (HRD) that shows the MESA-generated main sequence and subgiant branch of solitary stars with the masses of IT Lib's primary and secondary. The blue and red single points indicate the respective component's properties in the HRD. We see that the components do not land directly on the main sequence. The primary appears near the turn-off point between the main sequence and subgiant branch. The secondary, however, has a much higher luminosity than predicted for its mass, which points to an unusual evolutionary history for the secondary (Section \ref{sec:Orig}).

We next compare the evolutionary timescales to the time-of-flight from a starting point in the galactic plane to its current position in the halo. We calculated the trajectory using the galaxy modeling software, galpy \citep{2015ApJS..216...29B}. Galpy assumes a model for the gravitational potential and then uses the position and the velocities of a star system to integrate its motion through the Galaxy. The galactic potential used for this analysis was the one recommended for Milky Way modeling, ``MWPotential2014.'' This potential is described in the galpy documention \citep{2015ApJS..216...29B}. The coordinates and proper motion of IT Lib were taken from Gaia EDR3 \citep{2020yCat.1350....0G}. Its mean radial velocity $\gamma$ was set from the radial velocity solution (Table \ref{tab:paramfit}), and the distance was set to the result from the SED analysis (Section \ref{sec:SED}).

\begin{figure}[h!]
    \begin{center}
        \includegraphics[scale=0.8]{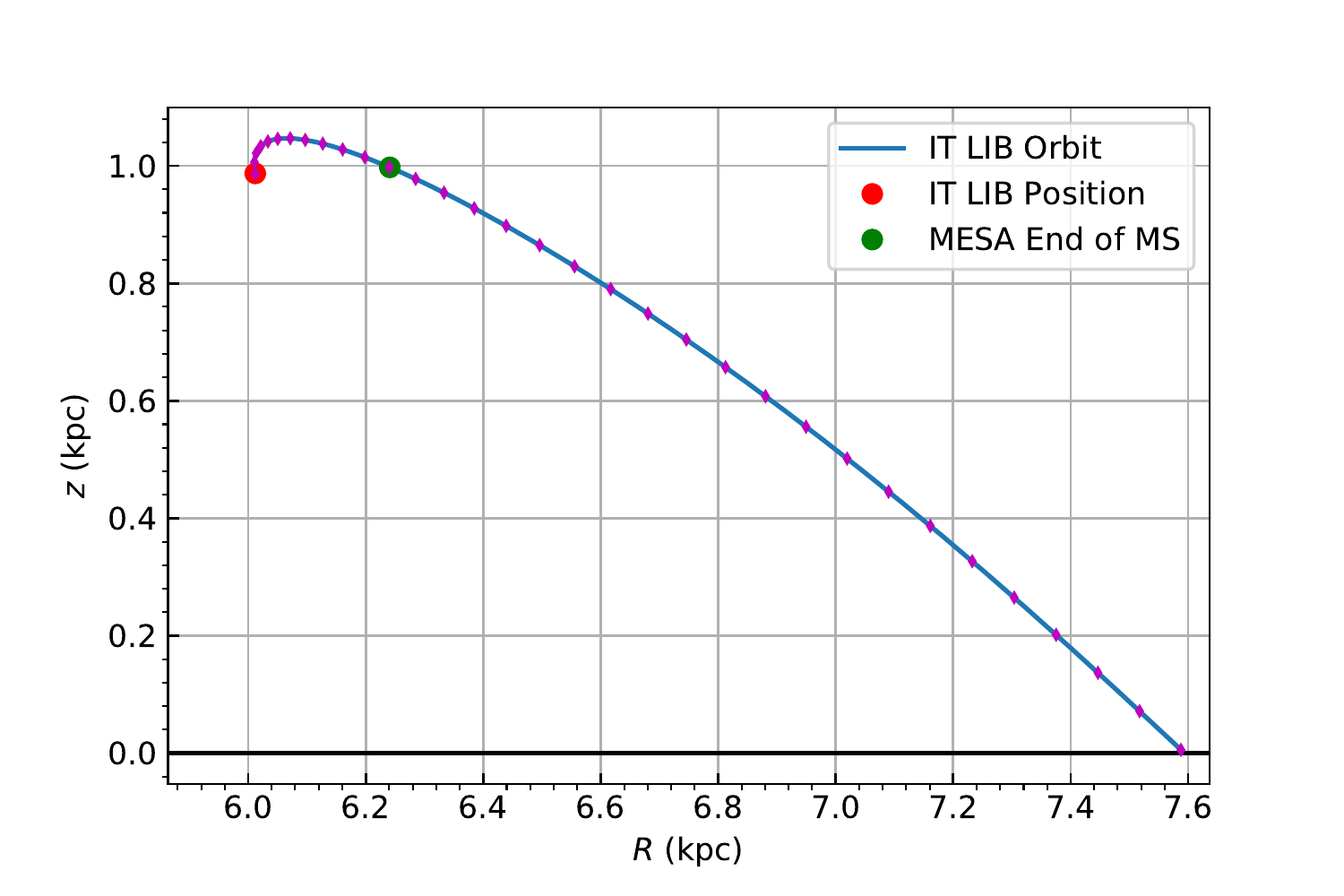}
    \caption{IT Lib's position over time in terms of height above the galactic plane $z$ and the distance from the Galactic center $R$. The solid black line marks the galactic plane. The red point is IT Lib's current position, the blue line shows positions where IT Lib was in the past, and the green point is where IT Lib's primary star should have left the main sequence according to MESA (if born around the time of ejection). The purple tick marks on the blue curve show 1 million year increments measured backwards from IT Librae's current position.}
    \label{fig:zR}
    \end{center}
\end{figure}

We then trace the orbit back through time from its current location to the galactic plane ($z=0$). This time-of-flight is 33 Myr, far exceeding the 21 Myr main sequence lifetime that is associated with a $9.6 M_{\odot}$ star like the primary of IT Lib. The calculated ejection velocity relative to the local standard of rest at the origin is 94 km~s$^{-1}$. Table \ref{tab:Travel} also lists the estimates of the time-of-flight and ejection velocity from the work of \cite{2003PASP..115...49M} and \cite{2011MNRAS.411.2596S} that agree broadly with our estimates. Figure \ref{fig:zR} shows IT Lib's position over time in terms of distance from the galactic mid-plane ($z$) and distance from the galactic center ($R$). Note that the trajectory indicates that IT Lib has already attained its greatest distance from the plane and is now returning towards the plane. This calculation verifies the discrepancy between the smaller evolutionary timescale and the longer time-of-flight.

We also attempted to identify the possible cluster of origin for IT Lib. From our travel time analysis, we know the location in the Galactic plane where IT Lib was located approximately 33 Myr ago. We then used the Galactic rotation curve for this position from galpy to find the current position of the origin environment assuming a circular motion around the galactic center for an elapsed time of 33 Myr and $z=0$.  The galactic latitude and distance to this current position is given in column 2 of Table \ref{tab:BirthCl}.  We then compared this position and age to the clusters listed in the catalog of \cite{2013AA...558A..53K}.  We found that Loden 821 (MWSC 2116) is the best candidate for IT Lib's birth cluster, and its properties are given in column 3 of Table \ref{tab:BirthCl}.  It should be noted that this analysis and identification carry significant uncertainties.  For example, there are uncertainties related to the original height of the 
birth cluster above or below the Galactic plane (along with its proper motion), the time elapsed between binary star birth and ejection, the details of the galactic potential (assumed cylindically symmetrical in galpy), and the observed cluster properties.  Nevertheless, the cluster Loden 821 shows the greatest consistency of properties for the origin of IT Lib than other nearby clusters.

Our kinematical results in Table \ref{tab:Travel} do not distinguish between the dynamical and supernova ejection processes for the origin of IT~Lib, but they do place some constraints on each.  In the dynamical model, the ejection of IT~Lib would be balanced by the release of a third star (or binary) in the 
opposite direction with the same momentum.  If the third star was as massive as the combined mass of IT~Lib, then it would have already ended its life in a supernova explosion and its remnant would be difficult to detect (particularly if the neutron star received a significant kick velocity).  A lower mass and high velocity third star could still exist, but tests with galpy indicate that it would probably be more than 10 kpc distant by now and too faint to be readily identified.  If, on the other hand, IT Lib was ejected by the supernova explosion of a tertiary star, then we can place some constraints on the tertiary mass and its separation from the close binary prior to the supernova.  We used the expression for the predicted ejection velocity from equation 7 in the work of \citealt{2019MNRAS.487.3178G} to estimate a solution set of tertiary mass $m_3$ and binary-tertiary semimajor axis $a_2$ given the derived ejection velocity (Table \ref{tab:Travel}) and assuming an initially circular orbit with a remnant mass $m_{\rm NS}=0.1 m_3$. The lower limit on $m_3$ is set by the condition that the binary-tertiary system is disrupted by the tertiary supernova ($v_{\rm rel} >0$ in equation 6 of \cite{2019MNRAS.487.3178G}), and this leads to a minimum tertiary mass of $m_3=17 M_\odot$ with a semimajor axis of 0.6 AU.  The derived ejection velocity leads to even larger separations at larger mass, i.e., 1.5 AU for $m_3 = 30 M_\odot$.  Thus, in the supernova scenario the tertiary would have been massive and short lived.

\begin{deluxetable*}{cccc}
\tablenum{6}
\tablecaption{IT Lib \textit{GalPy} Model \label{tab:Travel}}
\tablewidth{0pt}
\tablehead{
\colhead{Parameter} & \colhead{This Paper}& \colhead{Martin (2003)} & \colhead{Silva \& Napiwotzki (2011)}}
\decimalcolnumbers
\startdata
Distance (kpc)                          & 2.32       & 2.4            & 3.44    \\
$\mu_\alpha \cos \delta$ (mas yr$^{-1}$) & 1.52       & 1.82           & 0.61    \\
$\mu_\delta$ (mas yr$^{-1}$)             & 0.67       & $-1.26$        & $-0.39$ \\
$\gamma$ (km s$^{-1}$)                  & $-55$      & $-51$          & $-51$   \\
Time of flight (Myr)                    & 33         & 33             & 30      \\
$v_{\rm ejection}$ (km s$^{-1}$)        & 94         & \nodata        & 109     \\
\enddata
\end{deluxetable*}

\begin{deluxetable*}{ccc}
\tablenum{7}
\tablecaption{{\bf IT Lib Birth Cluster Potential Properties} \label{tab:BirthCl}}
\tablewidth{0pt}
\tablehead{
\colhead{Property} & \colhead{Target Cluster}& \colhead{Loden 821}}
\decimalcolnumbers
\startdata
Distance (kpc)                          & 2.76       & 2.64    \\
Galactic Latitude (deg) & 308.2       & 307.0      \\
Age (Myr)             & 33       & 29                  \\
Number of Associated Stars & \nodata & 277 \\
\enddata
\end{deluxetable*}

\section{System Origin} \label{sec:Orig}
Our results from the previous section indicate that the main sequence
lifetime of the primary is significantly less that the time-of-flight
if the binary was born in the vicinity of the galactic disk.
The solution to this dilemma may be that the primary began its
life as a lower mass and longer-lived star and that it was only
recently boosted in mass through mass transfer from its companion
(as proposed by \citealt{2009ApJ...698.1330P} for other apparently young objects
in the halo).  This solution is supported by two lines of evidence
concerning the secondary star.  First, the secondary fills its Roche
lobe (Section \ref{sec:ELC}), so it has the dimensions necessary to drive
mass transfer (now and in the past).  Second, the secondary is
overluminous for its mass compared to expectations for single stars
(Section \ref{sec:TTDet}), and this indicates its nuclear burning history is different
from that of similar mass, main sequence stars.

The current properties of IT Lib suggest that it is a post-mass transfer
binary that recently experienced a stage of large-scale mass transfer.
Close, short period, binary stars may begin interacting during the
slow expansion of the initially more massive component during
core hydrogen burning.  \citet{Wellstein2001} present an example
of the evolutionary progression of such Case~A mass transfer for
massive stars like those of IT~Lib. They show that the mass donor
experiences a rapid mass transfer stage that quickly leads to a
mass ratio reversal.  The donor enters a more extended slow mass
transfer stage (continuing to fill its Roche lobe) and appears as
a cooler and overluminous star.  The mass gainer star is boosted
to a higher mass position on the main sequence, and, with the
additional mass, it resets its evolutionary clock as a rejuvenated
star.

\begin{figure}[h!]
    \begin{center}
        \includegraphics[scale=0.8]{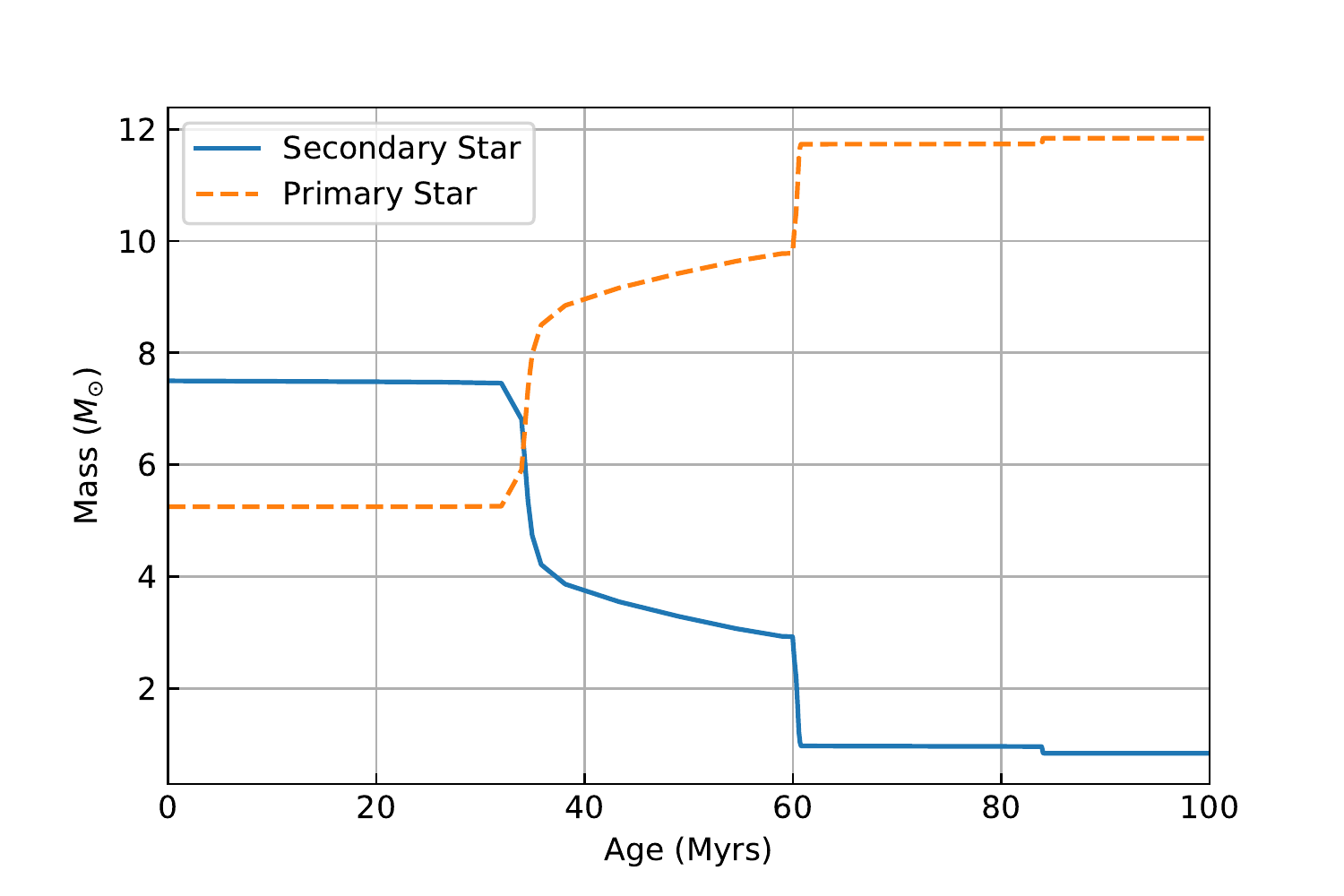}
    \caption{The changes in component masses over time in a BPASS model for Case A mass transfer (i.e., during core H-burning of the mass donor). This model predicts binary properties like those of IT Lib at an age of 36 Myr. The current secondary is the mass donor star in this scenario.}
    \label{fig:BPASS}
    \end{center}
\end{figure}

\cite{2017PASA...34...58E} have created a large collection of close
binary evolutionary tracks in the BPASS grid that are used primarily
for studying the influence of binaries for the properties of
stellar populations.  We inspected the BPASS sequences to find
examples that led to binary properties similar to those of IT~Lib.
The BPASS grid was calculated for a set of assumed values of
initial primary mass, mass ratio, and orbital period, and it is
helpful to use these published tracks for guidance about the
kinds of systems that pass through a stage like that of IT~Lib.
We formed a goodness-of-fit criterion by summing the squares
of the fractional differences (logarithms) between the observed
and model component temperatures, radii, masses, and orbital period
for each time step of the solar abundance models in BPASS.
Good matches were obtained with an initial primary mass in
the range 6 to $10 M_\odot$, initial mass ratio of 0.6 to 0.9
and initial period between 1.6 and 2.5 days.  The best fit was
obtained with a BPASS model with parameters given in Table \ref{tab:BPASS}.
Column 2 lists the starting values and column 3 gives the values
at an age of 36 Myr when these parameters are close to those
of IT Lib (column 4).  The main discrepancy is the smaller
predicted radius and lower predicted temperature of the present
day primary (mass gainer) star.  The variations in the model
component masses over time are shown in Figure \ref{fig:BPASS}.  The initially
more massive star (now the secondary) reaches its Roche lobe at an
age of about 32 Myr and it quickly transfers nearly half of
its mass to the gainer star (now the primary).  At 36 Myr,
the system enters a slower and extended mass transfer stage
and at that time it has many of the same properties found for IT~Lib.

If the evolutionary path of IT~Lib is similar to that in the
BPASS model, then the true age of the binary is probably
larger than the time-of-flight.  For example, if the actual
age is 36 Myr as found in the sample BPASS model, then the
binary may have resided in the disk for some 3 Myr before the
ejection event occurred (equal to the difference between the
true age and time-of-flight).  Thus, the interacting binary
scenario for IT~Lib offers an attractive explanation for
the apparent young age of the rejuvenated primary compared
to the time-of-flight from a position in the galactic plane.

\begin{deluxetable*}{cccc}
\tablenum{8}
\tablecaption{BPASS Fit to IT Lib \label{tab:BPASS}}
\tablewidth{0pt}
\tablehead{
\colhead{Parameter} & \colhead{BPASS Start} & \colhead{BPASS Current} & \colhead{IT Lib}}
\decimalcolnumbers
\startdata
Primary Mass ($M_{\odot}$)& 5.25 & 8.50 & 9.60\\
Secondary Mass ($M_{\odot}$)& 7.50 & 4.21 & 4.18 \\
Period (days)& 1.58 & 2.23 & 2.27 \\
Primary Temperature (kK)& 17.566 & 15.445 & 23.790 \\
Secondary Temperature (kK)& 18.684 & 13.326 & 13.680 \\
Primary Radius ($R_{\odot}$)& 2.71 & 3.91 & 6.06 \\
Secondary Radius ($R_{\odot}$)& 4.32 & 6.31 & 5.38 \\
\enddata
\end{deluxetable*}

\section{Discussion} \label{sec:Disc}

IT Lib is the nearest and brightest eclipsing binary known
among the high galactic latitude B-stars, and it is a key example to
better understand the nature of this unusual group of stars.
Our combined spectroscopic and photometric analysis of the system
provides reliable estimates for the stellar masses, radii, and
temperatures of the components. We find a secondary radius larger than was determined by previous work, owing to our analysis of shorter cadence and higher S/N photometry from Kepler K2 and our thorough exploration of the parameter space. For example, models with a near Roche-filling primary made a poor fit of the out-of-eclipse portions of the light curve. However, the models with a Roche filling secondary fit the entire light curve well and made a better match of the monochromatic flux ratio determined from spectroscopy.

A fit of the spectral energy
distribution based on the derived parameters yields a distance
that is in excellent agreement with that from Gaia EDR3.
The estimated evolutionary age for the primary of 21 Myr is
significantly smaller than the time-of-flight of 33 Myr if the
system was ejected from a birthplace near the Galactic plane.
This discrepancy in timescales is probably the result of a
binary interaction that has rejuvenated the primary, mass gainer
star to make it appear younger than it actually is.

We argue that IT Lib is in fact a post-mass transfer binary that
stripped the mass donor (now the secondary) of most of its envelope
and deposited this gas into the mass gainer (now the primary).
The secondary star fills its Roche lobe, has a much higher luminosity
than expected for its mass, and has He lines that are somewhat
stronger than predicted, perhaps indicating a He enhanced atmosphere.
These properties are consistent with predictions for a star
that has experienced stripping by mass transfer.  A comparison
to BPASS models for binary evolution suggests that the system
concluded large scale mass transfer relatively recently and
is now in a slow and long-lived mass transfer stage.  The actual
age of the system is much longer because the original mass donor
had a lower mass (and longer main sequence lifetime) than that of
the current primary star.  Thus, there was sufficient time for the
binary to reside in a typical massive star environment in the disk
before ejection to its current position in the halo.

We find marginal evidence of spectral line emission from on-going mass transfer,
and the H$\alpha$ feature is dominated by the absorption profiles of the
two stars.  This is not surprising given the expected lower mass transfer
rate at the current epoch and the small distance separating the stars.
\cite{1985PASP...97.1178K} showed that short-period Algol systems display
little H$\alpha$ emission, because the mass transfer gas stream directly
strikes the trailing hemisphere of the mass gainer rather than forming
a large accretion disk.  The photometric light curve presents an asymmetry
outside of eclipse that indicates greater flux from the trailing hemisphere
of the primary, and this may be due to local heating at the gas stream
impact site.

\cite{2009ApJ...698.1330P} presented a list of 16 runaway and hypervelocity stars
in the Galactic halo that appear too young, i.e., have an evolutionary
age less than the time-of-flight estimated for formation in the disk.
He argued that these systems could have extended lifetimes if they
were rejuvenated through mass transfer in a binary system.
The case of IT~Lib demonstrates that binary mass transfer can
transform the components and create mass gainer stars with short
apparent lifetimes.  Investigations of other halo massive stars
that appear too young may also reveal evidence of post-mass transfer
companions or binary mergers \citep{2022AJ....163..100G}.

\begin{acknowledgments}
This paper includes data collected by the Kepler mission and obtained from the MAST data archive at the Space Telescope Science Institute (STScI). Funding for the Kepler mission is provided by the NASA Science Mission Directorate. STScI is operated by the Association of Universities for Research in Astronomy, Inc., under NASA contract NAS 5-26555. Support for MAST for non-HST data is provided by the NASA Office of Space Science via grant NNX13AC07G and by other grants and contracts. We thank the New Mexico State University Department of Astronomy and the staff of the Apache Point Observatory. This work made use of v2.2.1 of the Binary Population and Spectral Synthesis (BPASS) models as described in \citet{2017PASA...34...58E} and \citet{Stanway2018}. This research has made use of the SIMBAD database, operated at CDS, Strasbourg, France. This research has used data from the CTIO/SMARTS 1.5m telescope, which is operated as part of the SMARTS Consortium by RECONS (\url{www.recons.org}) members Todd Henry, Hodari James, Wei-Chun Jao, and Leonardo Paredes. At the telescope, observations were carried out by Roberto Aviles and Rodrigo Hinojosa. This work has made use of data from the European Space Agency (ESA) mission {\it Gaia} (\url{https://www.cosmos.esa.int/gaia}), processed by the {\it Gaia} Data Processing and Analysis Consortium (DPAC, \url{https://www.cosmos.esa.int/web/gaia/dpac/consortium}). Funding for the DPAC has been provided by national institutions, in particular the institutions participating in the {\it Gaia} Multilateral Agreement.  The DASCH project is partially supported by NSF grants AST-0407380, AST-0909073, and AST-1313370. This work was supported in part by a Georgia State University Second Century Initiative (2CI) Fellowship to Peter Wysocki. The work was also supported by the National Science Foundation under Grant No. AST-1908026.
\end{acknowledgments}

\facilities{CTIO:1.5m, ARC:3.5m, VLT:Kueyen, Kepler}
\software{TLUSTY \citep{2007ApJS..169...83L}, rvfit.pro \citep{2015ascl.soft05020I}, BLUERED \citep{2008AA...485..823B}, ELC \citep{2000AA...364..265O}, TODCOR \citep{1994ApJ...420..806Z}, BPASS \citep{2017PASA...34...58E,Stanway2018}, MESA \citep{2015ApJS..220...15P,Fields2015}, galpy \citep{2015ApJS..216...29B}}

\bibliography{ITLIB2}{}
\bibliographystyle{aasjournal}



\end{document}